\documentclass[10pt]{article}
\usepackage[a4paper,margin=2.2cm]{geometry}
\usepackage[T1]{fontenc}
\usepackage[utf8]{inputenc}
\usepackage{lmodern}
\usepackage{microtype}
\usepackage{booktabs}
\usepackage{graphicx}
\usepackage{amsmath,amssymb}
\usepackage{xcolor}
\usepackage{listings}
\usepackage{caption}
\usepackage{subcaption}
\usepackage{enumitem}
\usepackage[hidelinks]{hyperref}
\usepackage{natbib}
\usepackage{pifont}

\newcommand{\nModels}{103}

\newcommand{\nHealthyRuns}{308}
\newcommand{\nHealthyFailRuns}{12}
\newcommand{\nHealthyFailModels}{4}

\newcommand{\nSideEffectRuns}{0}
\newcommand{\nDetPairs}{101}
\newcommand{\nDetIdentical}{100}

\newcommand{\detZooTT}{81}
\newcommand{\detZooTTN}{1411}
\newcommand{\detZooTTlenient}{90}

\newcommand{\detZooClassic}{61}

\newcommand{\detZooMonitor}{66}

\newcommand{\nMutFuzzModels}{600}
\newcommand{\detFuzzTT}{81}
\newcommand{\detFuzzTTN}{8935}
\newcommand{\detFuzzTTlenient}{99}

\newcommand{\detFuzzClassic}{52}

\newcommand{\detFuzzOverfit}{27}

\newcommand{\detFuzzSmoke}{7}

\newcommand{\faFuzzTT}{0/600}

\newcommand{\detHeldTT}{83}
\newcommand{\detHeldTTN}{2561}
\newcommand{\detHeldTTlenient}{91}

\newcommand{\detHeldClassic}{64}

\newcommand{\detHeldMonitor}{61}

\newcommand{\faHeldTT}{11/191}
\newcommand{\faHeldClassic}{45/191}

\newcommand{\faHeldMonitor}{24/191}

\newcommand{\pairZooTTPairedStrict}{80}
\newcommand{\pairZooTTPairedStrictLo}{79}
\newcommand{\pairZooTTPairedStrictHi}{82}

\newcommand{\pairZooTTPairedLenient}{91}
\newcommand{\pairZooTTPairedLenientLo}{89}
\newcommand{\pairZooTTPairedLenientHi}{92}

\newcommand{\pairZooTTReportStrict}{78}
\newcommand{\pairZooTTReportStrictLo}{74}
\newcommand{\pairZooTTReportStrictHi}{80}

\newcommand{\pairZooClassicPaired}{44}

\newcommand{\pairZooOverfitPaired}{24}

\newcommand{\pairZooSmokePaired}{3}

\newcommand{\pairZooMonitorPaired}{43}

\newcommand{\pairNoopZooTTLenient}{32/101}
\newcommand{\pairNoopZooTTPairedStrict}{0/101}
\newcommand{\pairNoopZooTTPairedLenient}{2/101}

\newcommand{\pairHealthyZooClassicRaw}{18}

\newcommand{\pairHealthyZooMonitorRaw}{23}

\newcommand{\pairDiffReportZooClassic}{34}
\newcommand{\pairDiffReportZooClassicLo}{29}
\newcommand{\pairDiffReportZooClassicHi}{38}

\newcommand{\pairDiffReportZooMonitor}{34}
\newcommand{\pairDiffReportZooMonitorLo}{28}
\newcommand{\pairDiffReportZooMonitorHi}{40}

\newcommand{\pairHeldTTPairedStrict}{81}
\newcommand{\pairHeldTTPairedStrictLo}{79}
\newcommand{\pairHeldTTPairedStrictHi}{82}

\newcommand{\pairHeldTTPairedLenient}{90}
\newcommand{\pairHeldTTPairedLenientLo}{89}
\newcommand{\pairHeldTTPairedLenientHi}{92}

\newcommand{\pairHeldTTReportStrict}{78}
\newcommand{\pairHeldTTReportStrictLo}{75}
\newcommand{\pairHeldTTReportStrictHi}{80}
\newcommand{\pairHeldTTReportStrictFamLo}{75}
\newcommand{\pairHeldTTReportStrictFamHi}{80}

\newcommand{\pairHeldClassicPaired}{40}

\newcommand{\pairHeldOverfitPaired}{22}

\newcommand{\pairHeldSmokePaired}{0}

\newcommand{\pairHeldMonitorPaired}{48}

\newcommand{\pairNoopHeldTTPairedLenient}{0/190}

\newcommand{\pairHealthyHeldTTLenient}{23}

\newcommand{\pairDiffStrictHeldClassic}{41}

\newcommand{\pairDiffStrictHeldMonitor}{32}

\newcommand{\pairDiffReportHeldClassic}{37}
\newcommand{\pairDiffReportHeldClassicLo}{34}
\newcommand{\pairDiffReportHeldClassicHi}{41}
\newcommand{\pairDiffReportHeldClassicFamLo}{33}
\newcommand{\pairDiffReportHeldClassicFamHi}{42}

\newcommand{\pairDiffReportHeldMonitor}{29}
\newcommand{\pairDiffReportHeldMonitorLo}{26}
\newcommand{\pairDiffReportHeldMonitorHi}{33}
\newcommand{\pairDiffReportHeldMonitorFamLo}{26}
\newcommand{\pairDiffReportHeldMonitorFamHi}{33}
\newcommand{\pairHeldFamilies}{120}

\newcommand{\pairFuzzTTPairedStrict}{81}

\newcommand{\pairFuzzTTPairedLenient}{99}

\newcommand{\costMedian}{1.0}
\newcommand{\costMax}{10}
\newcommand{\costRatio}{9}
\newcommand{\costOverfitShare}{52}
\newcommand{\stabGroups}{66}
\newcommand{\stabIdentical}{63}
\newcommand{\stabSameVerdict}{66}
\newcommand{\histFailRunsBefore}{63}
\newcommand{\histRunsBefore}{306}

\newcommand{\histFailModelsBefore}{21}

\newcommand{\devPairs}{135}
\newcommand{\devFailDiff}{0}
\newcommand{\devWarnDiff}{4}
\newcommand{\curN}{20}
\newcommand{\curTT}{18}

\newcommand{\curTTlenient}{20}

\newcommand{\curClassic}{8}

\newcommand{\curOverfit}{6}

\newcommand{\curSmoke}{1}

\newcommand{\scaleMaxParams}{91.4M}
\newcommand{\scaleMaxSeconds}{6}
\newcommand{\scaleMaxSecondsNoOverfit}{2.5}
\newcommand{\scaleMaxStep}{0.12}
\newcommand{\scaleMemRatio}{2.9}
\newcommand{\edgeCases}{41}

\newcommand{\edgeIssuesBefore}{10}
\newcommand{\rbTotal}{28}
\newcommand{\rbCrash}{13}
\newcommand{\rbNotBug}{6}
\newcommand{\rbOutScope}{6}
\newcommand{\rbInScope}{3}
\newcommand{\rvMainCompared}{1615}
\newcommand{\rvMainFailChanged}{2}
\newcommand{\rvMainStatusChanged}{26}

\newcommand{\rvHeldCompared}{2942}
\newcommand{\rvHeldFailChanged}{6}

\newcommand{\rvHeldPairedStrict}{81}
\newcommand{\rvHeldPairedLenient}{90}
\newcommand{\rvHeldReportStrict}{78}
\newcommand{\rvHeldHealthyFail}{10/191}

\newcommand{\rlMainCompared}{1615}
\newcommand{\rlMainFailChanged}{0}
\newcommand{\rlMainStatusChanged}{24}

\newcommand{\rlMainPairedStrict}{81}
\newcommand{\rlMainPairedLenient}{91}
\newcommand{\rlMainReportStrict}{79}

\newcommand{\rlHeldCompared}{2952}
\newcommand{\rlHeldFailChanged}{0}
\newcommand{\rlHeldStatusChanged}{5}

\newcommand{\rlHeldPairedStrict}{81}
\newcommand{\rlHeldPairedLenient}{90}
\newcommand{\rlHeldReportStrict}{78}

\newcommand{\rsMainCompared}{1615}
\newcommand{\rsMainFailChanged}{0}
\newcommand{\rsMainStatusChanged}{10}

\newcommand{\rsMainReportStrict}{79}

\newcommand{\rsHeldCompared}{2952}
\newcommand{\rsHeldFailChanged}{0}
\newcommand{\rsHeldStatusChanged}{8}

\newcommand{\rsHeldReportStrict}{78}

\newcommand{\costSevDefaultMedian}{0.8}
\newcommand{\costSevDefaultMax}{11}
\newcommand{\costSevDefaultBenchMedian}{0.8}
\newcommand{\costSevDefaultModels}{103}
\newcommand{\costSevDefaultOverfitWarn}{7}
\newcommand{\costSevDefaultBenchOverfitWarn}{10}
\newcommand{\tcTotal}{18}
\newcommand{\tcInScope}{3}
\newcommand{\tcPorted}{2}
\newcommand{\tcOutScope}{15}

\newcommand{\rbTT}{2}

\newcommand{\hoCandidates}{209}
\newcommand{\hoModels}{191}

\newcommand{\hoTunedFail}{3}

\newcommand{\hoHeldFail}{11}
\newcommand{\hoHeldFailPct}{6}

\newcommand{\hoHeldFixedN}{191}
\newcommand{\hoHeldFixedFail}{10}

\newcommand{\hoAllSeedsBefore}{14}
\newcommand{\hoAllSeedsAfter}{10}
\newcommand{\hoMainMutants}{1411}
\newcommand{\hoMainFailBefore}{1154}
\newcommand{\hoMainFailAfter}{1154}
\newcommand{\prN}{29}

\newcommand{\prMutants}{413}
\newcommand{\prDet}{86}
\newcommand{\prDetRandom}{79}
\newcommand{\pfCommit}{1dc4d58}
\newcommand{\pfN}{1411}
\newcommand{\pfStrictBefore}{81}
\newcommand{\pfStrictAfter}{82}
\newcommand{\pfLenientBefore}{90}
\newcommand{\pfLenientAfter}{91}
\newcommand{\pfHealthyFailBefore}{4}
\newcommand{\pfHealthyFailAfter}{3}
\newcommand{\pfHealthyN}{103}
\providecommand{\costMax}{\textbf{??}}
\providecommand{\costMedian}{\textbf{??}}
\providecommand{\costOverfitShare}{\textbf{??}}
\providecommand{\costRatio}{\textbf{??}}
\providecommand{\costSevDefaultBenchMedian}{\textbf{??}}
\providecommand{\costSevDefaultBenchOverfitWarn}{\textbf{??}}
\providecommand{\costSevDefaultMax}{\textbf{??}}
\providecommand{\costSevDefaultMedian}{\textbf{??}}
\providecommand{\costSevDefaultModels}{\textbf{??}}
\providecommand{\costSevDefaultOverfitWarn}{\textbf{??}}
\providecommand{\curClassic}{\textbf{??}}
\providecommand{\curN}{\textbf{??}}
\providecommand{\curOverfit}{\textbf{??}}
\providecommand{\curSmoke}{\textbf{??}}
\providecommand{\curTT}{\textbf{??}}
\providecommand{\curTTlenient}{\textbf{??}}
\providecommand{\detFuzzClassic}{\textbf{??}}
\providecommand{\detFuzzOverfit}{\textbf{??}}
\providecommand{\detFuzzSmoke}{\textbf{??}}
\providecommand{\detFuzzTT}{\textbf{??}}
\providecommand{\detFuzzTTN}{\textbf{??}}
\providecommand{\detFuzzTTlenient}{\textbf{??}}
\providecommand{\detHeldClassic}{\textbf{??}}
\providecommand{\detHeldMonitor}{\textbf{??}}
\providecommand{\detHeldTT}{\textbf{??}}
\providecommand{\detHeldTTN}{\textbf{??}}
\providecommand{\detHeldTTlenient}{\textbf{??}}
\providecommand{\detZooClassic}{\textbf{??}}
\providecommand{\detZooMonitor}{\textbf{??}}
\providecommand{\detZooTT}{\textbf{??}}
\providecommand{\detZooTTN}{\textbf{??}}
\providecommand{\detZooTTlenient}{\textbf{??}}
\providecommand{\devFailDiff}{\textbf{??}}
\providecommand{\devPairs}{\textbf{??}}
\providecommand{\devWarnDiff}{\textbf{??}}
\providecommand{\edgeCases}{\textbf{??}}
\providecommand{\edgeIssuesBefore}{\textbf{??}}
\providecommand{\faFuzzTT}{\textbf{??}}
\providecommand{\faHeldClassic}{\textbf{??}}
\providecommand{\faHeldMonitor}{\textbf{??}}
\providecommand{\faHeldTT}{\textbf{??}}
\providecommand{\histFailModelsBefore}{\textbf{??}}
\providecommand{\histFailRunsBefore}{\textbf{??}}
\providecommand{\histRunsBefore}{\textbf{??}}
\providecommand{\hoAllSeedsAfter}{\textbf{??}}
\providecommand{\hoAllSeedsBefore}{\textbf{??}}
\providecommand{\hoCandidates}{\textbf{??}}
\providecommand{\hoHeldFail}{\textbf{??}}
\providecommand{\hoHeldFailPct}{\textbf{??}}
\providecommand{\hoHeldFixedFail}{\textbf{??}}
\providecommand{\hoHeldFixedN}{\textbf{??}}
\providecommand{\hoMainFailAfter}{\textbf{??}}
\providecommand{\hoMainFailBefore}{\textbf{??}}
\providecommand{\hoMainMutants}{\textbf{??}}
\providecommand{\hoModels}{\textbf{??}}
\providecommand{\hoTunedFail}{\textbf{??}}
\providecommand{\nDetIdentical}{\textbf{??}}
\providecommand{\nDetPairs}{\textbf{??}}
\providecommand{\nHealthyFailModels}{\textbf{??}}
\providecommand{\nHealthyFailRuns}{\textbf{??}}
\providecommand{\nHealthyRuns}{\textbf{??}}
\providecommand{\nModels}{\textbf{??}}
\providecommand{\nMutFuzzModels}{\textbf{??}}
\providecommand{\nSideEffectRuns}{\textbf{??}}
\providecommand{\pairDiffReportHeldClassic}{\textbf{??}}
\providecommand{\pairDiffReportHeldClassicFamHi}{\textbf{??}}
\providecommand{\pairDiffReportHeldClassicFamLo}{\textbf{??}}
\providecommand{\pairDiffReportHeldClassicHi}{\textbf{??}}
\providecommand{\pairDiffReportHeldClassicLo}{\textbf{??}}
\providecommand{\pairDiffReportHeldMonitor}{\textbf{??}}
\providecommand{\pairDiffReportHeldMonitorFamHi}{\textbf{??}}
\providecommand{\pairDiffReportHeldMonitorFamLo}{\textbf{??}}
\providecommand{\pairDiffReportHeldMonitorHi}{\textbf{??}}
\providecommand{\pairDiffReportHeldMonitorLo}{\textbf{??}}
\providecommand{\pairDiffReportZooClassic}{\textbf{??}}
\providecommand{\pairDiffReportZooClassicHi}{\textbf{??}}
\providecommand{\pairDiffReportZooClassicLo}{\textbf{??}}
\providecommand{\pairDiffReportZooMonitor}{\textbf{??}}
\providecommand{\pairDiffReportZooMonitorHi}{\textbf{??}}
\providecommand{\pairDiffReportZooMonitorLo}{\textbf{??}}
\providecommand{\pairDiffStrictHeldClassic}{\textbf{??}}
\providecommand{\pairDiffStrictHeldMonitor}{\textbf{??}}
\providecommand{\pairFuzzTTPairedLenient}{\textbf{??}}
\providecommand{\pairFuzzTTPairedStrict}{\textbf{??}}
\providecommand{\pairHealthyHeldTTLenient}{\textbf{??}}
\providecommand{\pairHealthyZooClassicRaw}{\textbf{??}}
\providecommand{\pairHealthyZooMonitorRaw}{\textbf{??}}
\providecommand{\pairHeldClassicPaired}{\textbf{??}}
\providecommand{\pairHeldFamilies}{\textbf{??}}
\providecommand{\pairHeldMonitorPaired}{\textbf{??}}
\providecommand{\pairHeldOverfitPaired}{\textbf{??}}
\providecommand{\pairHeldSmokePaired}{\textbf{??}}
\providecommand{\pairHeldTTPairedLenient}{\textbf{??}}
\providecommand{\pairHeldTTPairedLenientHi}{\textbf{??}}
\providecommand{\pairHeldTTPairedLenientLo}{\textbf{??}}
\providecommand{\pairHeldTTPairedStrict}{\textbf{??}}
\providecommand{\pairHeldTTPairedStrictHi}{\textbf{??}}
\providecommand{\pairHeldTTPairedStrictLo}{\textbf{??}}
\providecommand{\pairHeldTTReportStrict}{\textbf{??}}
\providecommand{\pairHeldTTReportStrictFamHi}{\textbf{??}}
\providecommand{\pairHeldTTReportStrictFamLo}{\textbf{??}}
\providecommand{\pairHeldTTReportStrictHi}{\textbf{??}}
\providecommand{\pairHeldTTReportStrictLo}{\textbf{??}}
\providecommand{\pairNoopHeldTTPairedLenient}{\textbf{??}}
\providecommand{\pairNoopZooTTLenient}{\textbf{??}}
\providecommand{\pairNoopZooTTPairedLenient}{\textbf{??}}
\providecommand{\pairNoopZooTTPairedStrict}{\textbf{??}}
\providecommand{\pairZooClassicPaired}{\textbf{??}}
\providecommand{\pairZooMonitorPaired}{\textbf{??}}
\providecommand{\pairZooOverfitPaired}{\textbf{??}}
\providecommand{\pairZooSmokePaired}{\textbf{??}}
\providecommand{\pairZooTTPairedLenient}{\textbf{??}}
\providecommand{\pairZooTTPairedLenientHi}{\textbf{??}}
\providecommand{\pairZooTTPairedLenientLo}{\textbf{??}}
\providecommand{\pairZooTTPairedStrict}{\textbf{??}}
\providecommand{\pairZooTTPairedStrictHi}{\textbf{??}}
\providecommand{\pairZooTTPairedStrictLo}{\textbf{??}}
\providecommand{\pairZooTTReportStrict}{\textbf{??}}
\providecommand{\pairZooTTReportStrictHi}{\textbf{??}}
\providecommand{\pairZooTTReportStrictLo}{\textbf{??}}
\providecommand{\pfCommit}{\textbf{??}}
\providecommand{\pfHealthyFailAfter}{\textbf{??}}
\providecommand{\pfHealthyFailBefore}{\textbf{??}}
\providecommand{\pfHealthyN}{\textbf{??}}
\providecommand{\pfLenientAfter}{\textbf{??}}
\providecommand{\pfLenientBefore}{\textbf{??}}
\providecommand{\pfN}{\textbf{??}}
\providecommand{\pfStrictAfter}{\textbf{??}}
\providecommand{\pfStrictBefore}{\textbf{??}}
\providecommand{\prDet}{\textbf{??}}
\providecommand{\prDetRandom}{\textbf{??}}
\providecommand{\prMutants}{\textbf{??}}
\providecommand{\prN}{\textbf{??}}
\providecommand{\rbCrash}{\textbf{??}}
\providecommand{\rbInScope}{\textbf{??}}
\providecommand{\rbNotBug}{\textbf{??}}
\providecommand{\rbOutScope}{\textbf{??}}
\providecommand{\rbTT}{\textbf{??}}
\providecommand{\rbTotal}{\textbf{??}}
\providecommand{\rlHeldCompared}{\textbf{??}}
\providecommand{\rlHeldFailChanged}{\textbf{??}}
\providecommand{\rlHeldPairedLenient}{\textbf{??}}
\providecommand{\rlHeldPairedStrict}{\textbf{??}}
\providecommand{\rlHeldReportStrict}{\textbf{??}}
\providecommand{\rlHeldStatusChanged}{\textbf{??}}
\providecommand{\rlMainCompared}{\textbf{??}}
\providecommand{\rlMainFailChanged}{\textbf{??}}
\providecommand{\rlMainPairedLenient}{\textbf{??}}
\providecommand{\rlMainPairedStrict}{\textbf{??}}
\providecommand{\rlMainReportStrict}{\textbf{??}}
\providecommand{\rlMainStatusChanged}{\textbf{??}}
\providecommand{\rsHeldCompared}{\textbf{??}}
\providecommand{\rsHeldFailChanged}{\textbf{??}}
\providecommand{\rsHeldReportStrict}{\textbf{??}}
\providecommand{\rsHeldStatusChanged}{\textbf{??}}
\providecommand{\rsMainCompared}{\textbf{??}}
\providecommand{\rsMainFailChanged}{\textbf{??}}
\providecommand{\rsMainReportStrict}{\textbf{??}}
\providecommand{\rsMainStatusChanged}{\textbf{??}}
\providecommand{\rvHeldCompared}{\textbf{??}}
\providecommand{\rvHeldFailChanged}{\textbf{??}}
\providecommand{\rvHeldHealthyFail}{\textbf{??}}
\providecommand{\rvHeldPairedLenient}{\textbf{??}}
\providecommand{\rvHeldPairedStrict}{\textbf{??}}
\providecommand{\rvHeldReportStrict}{\textbf{??}}
\providecommand{\rvMainCompared}{\textbf{??}}
\providecommand{\rvMainFailChanged}{\textbf{??}}
\providecommand{\rvMainStatusChanged}{\textbf{??}}
\providecommand{\scaleMaxParams}{\textbf{??}}
\providecommand{\scaleMaxSeconds}{\textbf{??}}
\providecommand{\scaleMaxSecondsNoOverfit}{\textbf{??}}
\providecommand{\scaleMaxStep}{\textbf{??}}
\providecommand{\scaleMemRatio}{\textbf{??}}
\providecommand{\stabGroups}{\textbf{??}}
\providecommand{\stabIdentical}{\textbf{??}}
\providecommand{\stabSameVerdict}{\textbf{??}}
\providecommand{\tcInScope}{\textbf{??}}
\providecommand{\tcOutScope}{\textbf{??}}
\providecommand{\tcPorted}{\textbf{??}}
\providecommand{\tcTotal}{\textbf{??}}
\graphicspath{{figures/}}

\definecolor{codebg}{HTML}{F6F6F4}
\newcommand{\tool}{\textsf{torchtest2}}
\newcommand{\chk}[1]{\texttt{#1}}
\newcommand{\RQ}[1]{\textbf{RQ#1}}
\setlist[itemize]{leftmargin=*,itemsep=1pt}

\title{\tool: One-Call Sanity Checks for PyTorch Models,\\
Evaluated on Real and Unseen Architectures}
\author{%
  \begin{tabular}{c}
    Grzegorz Gwardys$^{1,2}$\\[4pt]
    $^{1}$Faculty of Electronics and Information Technology,\\ Warsaw University of Technology,\\
    Warsaw, Poland\\
    \texttt{grzegorz.gwardys@pw.edu.pl}\\[4pt]
    $^{2}$Promity\\
    \texttt{grzegorz.gwardys@promity.pl}
  \end{tabular}}
\date{October 2026}

\begin{document}
\maketitle

\begin{abstract}
Many bugs in deep learning code are \emph{silent}: a layer that is never called, an optimizer
created before the classification head was replaced, a recurrent layer that treats the batch as
time, or an attention mask that is ignored. The code runs, the loss decreases somewhat, and the
defect surfaces---if ever---after days of GPU time. We present \tool{}, an open-source library
that runs 18 sanity checks on a single batch with one call, restores the PyTorch state it touches
(parameters, buffers, optimizer state, gradients, train/eval flags, random generators) and adapts
to model families through plugins. We evaluate it by mutation analysis with 17 classes of silent
bugs on \nModels{} healthy models from torchvision, timm and Hugging Face transformers, on 600
random architectures and on \curN{} hand-written realistic bugs, against the checks of the
original torchtest, overfitting a single batch, a smoke test and a training monitor in the style
of AUTOTRAINER and DeepDiagnosis. Every mutant is compared with its healthy counterpart, so that
an alarm the unmodified model already raises does not count, and intervals come from a bootstrap
over models. Because the fixes that the evaluation prompted were tuned on the same zoo, we froze
the tool and ran it on \hoModels{} models of architecture families it had never seen. Counting,
for every tool alike, a failing report where the healthy model's report does not fail, it
detected \pairHeldTTReportStrict\% of the injected bugs (95\% CI
\pairHeldTTReportStrictLo--\pairHeldTTReportStrictHi), against \pairHeldClassicPaired\% for the
original torchtest checks and \pairHeldMonitorPaired\% for the training monitor, and no
equivalent mutant; its per-check report shows a new problem for \pairHeldTTPairedLenient\% of
the bugs (\pairHeldTTPairedStrict\% with a new \texttt{FAIL}).
On real models its batch-independence check found a bug in timm's GCViT that makes predictions
depend on the other images in a batch, and Hugging Face models that ignore their attention mask.
Evaluating the tool on real models was essential: it exposed false alarms and robustness defects
that neither its own test suite nor random architectures had revealed. Two independent
collections of real silent bugs mark the limits of the approach: of \rbTotal{} reproduced
PyTorch bugs from Stack Overflow only \rbInScope{}, and of the \tcTotal{} silent training errors
in TrainCheck's public artefact only \tcInScope, are within reach of a pre-training model tester;
\tool{} flags all those we could port. A median check takes \costMedian\,s. We release the
library, the benchmark and all raw results; the replication package is archived on Zenodo
(\url{https://doi.org/10.5281/zenodo.23301395}).
\end{abstract}

\section{Introduction}

Testing deep learning (DL) code is hard for a reason that is independent of model quality: the
code under test is a numerical program whose correct output is unknown. Studies of real DL faults
\citep{zhang2018tfbugs,islam2019bugs,humbatova2020taxonomy} consistently find that a large share
of them does not crash the program. A layer that is defined but never used, a computation graph
cut by \texttt{.detach()}, a batch dimension confused with a time dimension, or an optimizer that
does not own all parameters, all produce a model that trains---just worse than it should.

Practitioners fight these bugs with folklore: ``overfit a single batch'', ``check the loss at
initialisation'' \citep{karpathy2019recipe}, ``check that all variables change''
\citep{roberts2017mltest}, ``check that samples do not influence each other''
\citep{krokotsch2020unittests}. Small libraries packaged some of this folklore
\citep{roberts2017mltest,torchtest2019}, but they are unmaintained, cover four checks, and modify
the model they test. Research tools detect training problems by monitoring training runs
\citep{braiek2019tfcheck,zhang2021autotrainer,wardat2022deepdiagnosis,braiek2023thedeepchecker}
or, as TrainCheck \citep{jiang2025traincheck}, by checking a run against invariants inferred
from traces of correct reference pipelines. Both need a training run---and the latter a known-good
one---which is precisely what is missing while a new architecture is being written.

We present \tool{}, a PyTorch \citep{paszke2019pytorch} library designed around one call:

\begin{lstlisting}
report = torchtest2.check_model(model, inputs, targets,
                                loss_fn=F.cross_entropy, optimizer=optimizer)
print(report)   # 18 checks: PASS / WARN / FAIL / SKIP, each with an explanation
\end{lstlisting}

\noindent and evaluate it empirically. Our contributions are:
\begin{itemize}
  \item A set of 18 checks (Section~\ref{sec:design}) that each target a class of silent bug,
    including two checks that we believe are new in this form: \emph{batch independence by
    sample substitution}, which, unlike gradient-based formulations, needs neither differentiable
    inputs nor a per-sample loss, and \emph{input dependency}
    that combines per-feature gradients with value-preserving perturbations of discrete inputs.
  \item An execution model that restores the PyTorch training state it touches---parameters,
    buffers, optimizer state, gradients, train/eval flags and the PyTorch RNG---and copies the
    inputs, and a plugin mechanism that adapts the checks to model families whose training step
    is not \texttt{model(inputs)} (detection models, Hugging Face models, Lightning modules).
  \item A benchmark: \nModels{} models from three major model libraries, 17 mutation operators
    for silent bugs plus an equivalent-mutant control, four baselines, a fuzzing study on random
    architectures, an edge-case study, and measurements of cost, determinism and stability.
    Detection is counted relative to each mutant's healthy counterpart, with confidence
    intervals from a bootstrap over models and paired comparisons between tools.
  \item A held-out validation on \hoModels{} models of unseen architecture families, selected and
    committed before any run, an evaluation on pretrained weights, a training-monitor baseline,
    and a scope analysis of real silent bugs from two independent sources: a dataset of
    reproduced Stack Overflow bugs and the silent training errors released in TrainCheck's public
    artefact.
  \item Empirical findings: a real batch-coupling bug in timm's GCViT; genuinely unused
    parameters and over-confident default initialisations in widely used implementations; and
    the defects the evaluation revealed in \tool{} itself---a reminder that a model tester has
    to be validated on real models, not only on synthetic ones.
\end{itemize}

\section{Background and related work}
\label{sec:related}

\paragraph{Faults in DL programs.} Empirical studies of bugs in TensorFlow and PyTorch programs
\citep{zhang2018tfbugs,islam2019bugs} and the taxonomy of \citet{humbatova2020taxonomy} group
faults into model (layers, activations), tensor and input, training (loss, optimizer,
hyper-parameters) and GPU usage. Many categories in these studies correspond to defects that
produce no error message. Surveys of ML testing \citep{zhang2020mltesting,riccio2020mapping,
braiek2020testing} and production rubrics \citep{breck2017mltestscore,sculley2015debt} note the
scarcity of practical, automated unit-level checks.

\paragraph{Detecting training problems.} TFCheck \citep{braiek2019tfcheck}, AUTOTRAINER
\citep{zhang2021autotrainer}, DeepLocalize \citep{wardat2021deeplocalize} and DeepDiagnosis
\citep{wardat2022deepdiagnosis} monitor quantities such as activations, gradients and the loss
during training to detect and sometimes repair problems like vanishing gradients or dying ReLUs.
TheDeepChecker \citep{braiek2023thedeepchecker}, a TensorFlow library, is the closest in spirit
to \tool: it verifies properties of feed-forward training programs in three phases---conditions
before training (input data, initial weights, the expected initial loss), fitting a single batch,
and the dynamics of the on-going training---and it was evaluated on synthetic and real buggy
programs. TrainCheck
\citep{jiang2025traincheck} instruments training pipelines, infers invariants of API calls and
variable states from traces of known-good pipelines, and checks new runs against them; it
detected 18 of 20 real silent errors, most of them in distributed training, compilers and
frameworks, within one iteration. \tool{} differs in its inputs and in where it runs: it needs
no reference run and no instrumentation, only a model and one batch, its invariants are
hand-written and model-level (gradients, parameter updates, coupling between samples, input
influence), and it is meant to run as a unit test while the model is being written, before any
training pipeline exists.
UMLAUT \citep{schoop2021umlaut} combines static program structure, runtime behaviour and error
messages into heuristics shown to users; NeuraLint \citep{nikanjam2021neuralint} detects faults
statically with graph transformations. Cockpit \citep{schneider2021cockpit} instruments PyTorch
training with gradient and curvature statistics, but leaves their interpretation to the user: it
gives no automatic verdict, so it cannot serve as a baseline without inventing decision rules. \tool{} is complementary: it targets defects that are
visible \emph{before} training, from a single batch, in seconds, and it is designed to run in a
unit-test suite.

\paragraph{Unit-test libraries.} mltest \citep{roberts2017mltest} (TensorFlow) and torchtest
\citep{torchtest2019} (PyTorch) check that variables change after a training step, that outputs
are finite and within a range, and that a GPU is used. Both perform the training step on the
model under test and leave it modified. Lightning's \texttt{fast\_dev\_run} and
\texttt{overfit\_batches} \citep{falcon2019lightning} run the training loop quickly but leave the
diagnosis to the user.

\paragraph{Validating trained models.} Deepchecks \citep{chorev2022deepchecks} and Giskard
\citep{giskard} test data and \emph{trained} models (drift, leakage, performance, robustness);
experiment trackers plot gradient histograms. They work after training, \tool{} before it.

\paragraph{Datasets of DL bugs.} \citet{shah2024reproducibility} reproduced 148 bugs from Stack
Overflow and GitHub; \citet{jahangirova2024realfaults} found that only a minority of the faults
in five DL fault benchmarks satisfy their realism conditions, a caution for any evaluation built
on injected faults, including ours.

\paragraph{Mutation testing of DL.} DeepMutation \citep{ma2018deepmutation} and DeepCrime
\citep{humbatova2021deepcrime} mutate DL programs or models to assess test \emph{data}; DeepCrime
derives its operators from real faults. We use mutation analysis \citep{jia2011mutation} for the
opposite purpose: to assess a \emph{test tool}, injecting silent program bugs into healthy models
and measuring which tools notice them.

\paragraph{Behavioural testing.} CheckList \citep{ribeiro2020checklist} introduced minimum
functionality, invariance and directional tests for NLP models; \tool{} ships these three test
types for trained models, but this paper focuses on the pre-training checks.

\section{Design of \tool}
\label{sec:design}

\subsection{Requirements}
The design follows from how the tool is meant to be used---inside a unit test, while an
architecture is being written:
\textbf{(R1)} one call with sensible defaults;
\textbf{(R2)} no effect on the PyTorch state the user passes in, so that it can run anywhere in a
test suite;
\textbf{(R3)} deterministic reports;
\textbf{(R4)} unmodified \texttt{nn.Module}s whose inputs and outputs are tensors nested in tuples,
lists, dicts, namedtuples, dataclasses or Hugging Face output objects, with plugins for families
that are trained differently;
\textbf{(R5)} actionable messages that name the offending parameters and suggest likely causes;
\textbf{(R6)} a severity model that separates near-certain bugs (\texttt{FAIL}) from suspicious
but possibly legitimate findings (\texttt{WARN}).

\subsection{Checks}
Table~\ref{tab:checks} lists the checks in execution order. Most are simple individually; the
value lies in running them together, consistently, with a common execution model. We describe
the less obvious ones.

\begin{table}[htbp]
\centering\footnotesize
\caption{The checks of \tool, in execution order. \emph{Mode}: eval or train forward passes.}
\label{tab:checks}
\begin{tabular}{llp{0.5\linewidth}l}
\toprule
Check & Mode & Verifies & Severity \\
\midrule
\chk{forward} & eval & the model runs on the batch & FAIL \\
\chk{output\_shape} & eval & output shape matches the expectation (optional) & FAIL \\
\chk{output\_finite} & eval & no NaN/inf in any output & FAIL \\
\chk{output\_range} & eval & outputs within bounds (optional) & FAIL \\
\chk{output\_varies} & eval & not every sample yields the same output & WARN \\
\chk{eval\_determinism} & eval & two eval passes agree (no randomness at inference) & WARN \\
\chk{loss} & train & finite scalar, connected to the graph, not a re-created leaf & FAIL \\
\chk{initial\_loss} & train & cross-entropy at init close to $\ln C$ & WARN \\
\chk{optimizer\_coverage} & -- & every trainable parameter is in the optimizer & FAIL \\
\chk{gradient\_flow} & train & every trainable parameter receives a gradient & FAIL/WARN \\
\chk{gradient\_magnitude} & train & gradients finite; not large relative to the parameter & FAIL/WARN \\
\chk{params\_change} & train & one optimizer step changes every trainable parameter & FAIL/WARN \\
\chk{frozen\_params} & train & frozen parameters stay fixed & FAIL \\
\chk{input\_dependency} & eval & every input and every input feature influences the output & FAIL/WARN \\
\chk{batch\_independence} & eval & a sample's output does not depend on other samples & FAIL/WARN \\
\chk{overfit\_batch} & train & loss drops on one batch, gradients stay finite & FAIL/WARN \\
\chk{gradcheck} & eval & analytical = numerical gradients, float64 copy (optional) & FAIL \\
\chk{gpu} & -- & CUDA available and used (optional) & FAIL \\
\bottomrule
\end{tabular}
\end{table}

\paragraph{Surrogate objective.} Gradient-based checks need a scalar objective even when the user
gives no loss. The natural choice, the sum of all outputs, is constant for normalised outputs
($\sum_i \mathrm{softmax}(z)_i = 1$), which would make every parameter look disconnected. \tool{}
instead uses a fixed random projection $\sum_k \langle o_k, w_k\rangle$ with $w_k \sim
\mathcal{N}(0, I)$ over all floating-point (and complex) outputs $o_k$. Whenever the Jacobian of
the outputs with respect to a parameter is non-zero at the evaluated point, the gradient of this
projection is non-zero almost surely (a parameter can influence the outputs and still have a zero
local Jacobian, e.g.\ $f(\theta)=\theta^2$ at $\theta=0$).

\paragraph{Batch independence by sample substitution.} \citet{krokotsch2020unittests} proposed
masking one sample's loss and checking that only its input gradient vanishes. This requires
differentiable inputs and a per-sample loss structure. \tool{} instead replaces sample~0 of every
batched input by sample~1 and checks, in eval mode, that the outputs of samples $1\ldots N-1$ are
unchanged. This works for integer inputs (token ids), boolean masks, and nested tensor outputs
(tuples, lists, dicts, dataclasses and Hugging Face outputs), given a batch dimension,
and catches recurrent or attention layers used with the wrong \texttt{batch\_first}, reshapes
that mix samples, normalisation over the batch dimension, and batch statistics at inference.

\paragraph{Input dependency.} For floating-point inputs, \tool{} computes the gradient of the
surrogate objective with respect to each input and reports features whose gradient vanishes for
every sample. Features that are exactly zero in every sample are exempt, because they cannot be
distinguished from padding. Integer and boolean inputs cannot be differentiated; they are rolled
along the batch dimension, a perturbation that keeps values valid, and the check reports inputs
whose perturbation leaves all outputs unchanged. A perturbation that the model rejects (e.g.\ it
requires sorted lengths) is reported as a note instead of a failure. Two fallbacks avoid false
alarms that we observed on real models (Section~\ref{sec:rq1}): features without influence on the
eval outputs are re-checked through the training objective (detection outputs are piecewise
constant after non-maximum suppression) and after one optimizer step (zero-initialised heads and
residual branches hide inputs at initialisation). Features missed only at the trailing border of
a spatial input---pixels a strided convolution never covers---give a warning.

\paragraph{Parameter checks.} \chk{gradient\_flow} distinguishes parameters with no gradient
(\texttt{None}: unused, or the graph is cut; \texttt{FAIL}) from parameters with an all-zero
gradient (\texttt{WARN}); the latter is common and intended with zero-initialised layers.
\chk{params\_change} performs one step with the user's optimizer, supporting optimizers that
require a closure, and compares with the snapshot; a parameter that did not change because its
gradient was exactly zero is a warning, one that did not change despite a non-zero gradient an
optimizer failure. \chk{gradient\_magnitude} relates each gradient norm to the norm of its
parameter, and \chk{initial\_loss} measures the loss in training mode---in eval mode, the
running statistics of untrained BatchNorm layers inflate the logits of deep networks.
\chk{optimizer\_coverage} compares parameter identities with the optimizer's parameter groups
and recognises optimizers created for another model instance.

\paragraph{Overfitting.} \chk{overfit\_batch} trains on the batch until the loss drops to a
fraction (default 10\%) of its initial value or an absolute target, and fails if the loss
becomes non-finite or does not decrease. It also checks gradient finiteness at \emph{every}
step, which catches NaN gradients that only appear after a few updates; an option runs the loop
under \texttt{torch.autograd.detect\_anomaly} to name the offending operation.

\subsection{Execution model}
\label{sec:execution}
\begin{figure}[htbp]
\centering
\includegraphics[width=\linewidth]{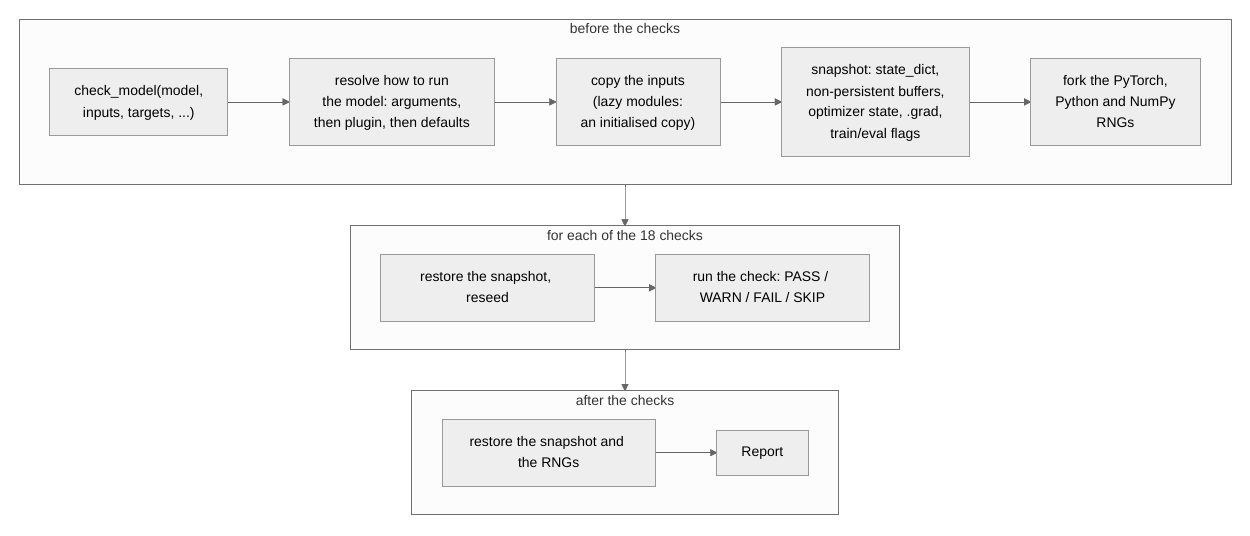}
\caption{What \texttt{check\_model} does around the checks: every check starts from the same
snapshot with freshly seeded random generators, and the PyTorch state is restored at the end.}
\label{fig:lifecycle}
\end{figure}

Figure~\ref{fig:lifecycle} summarises a call. Before the checks run, \tool{} snapshots the model's \texttt{state\_dict} (on the CPU) and the
non-persistent buffers it leaves out, a deep copy of the optimizer state, existing \texttt{.grad} tensors and every module's train/eval flag,
and restores them before each check and at the end. The checks run inside
\texttt{torch.random.fork\_rng} and reseed it, so the PyTorch CPU and CUDA generators are
restored afterwards and reports are deterministic. Forward passes whose results are compared with
each other are reseeded identically, so that randomness at inference (reported by
\chk{eval\_determinism}) does not cause false alarms in other checks. Every tensor in the inputs
is copied, so models that modify their inputs in place cannot affect the caller. Outputs of two
passes are compared only after checking that they have the same structure (the same tensors under
the same keys, with the same shapes and dtypes); a change of structure is itself reported.

The guarantee is deliberately limited to what PyTorch exposes: plain Python attributes that the
model's own code changes (a call counter, a cache) are not restored, nor are tensors held in
objects that \tool{} does not traverse. A review of an earlier version of this paper found three
gaps in V1--V4 that the side-effect fingerprint of our benchmark could not see,
because no benchmark model exercised them: tensors in dataclass inputs were not copied, lazy
modules stayed initialised after the call, and the Python and NumPy random generators were
neither seeded nor restored. The current version closes them: dataclasses are copied field by
field, a model with uninitialised lazy modules is checked as an initialised copy (from V6 on
together with a copy of the optimizer when one is passed; V5 initialised the caller's modules in
that case), and the Python and NumPy global generators are seeded for each check and restored at
the end. A third review found one more gap, closed in V7: buffers registered with
\texttt{persistent=False}, which \texttt{state\_dict()} leaves out, were not restored. Re-running
both zoos with each of these versions changed
only verdicts that these fixes were meant to change or that are known to be unstable
(Section~\ref{sec:heldoutfixes}).

\subsection{Plugins}
\begin{figure}[htbp]
\centering
\includegraphics[width=0.7\linewidth]{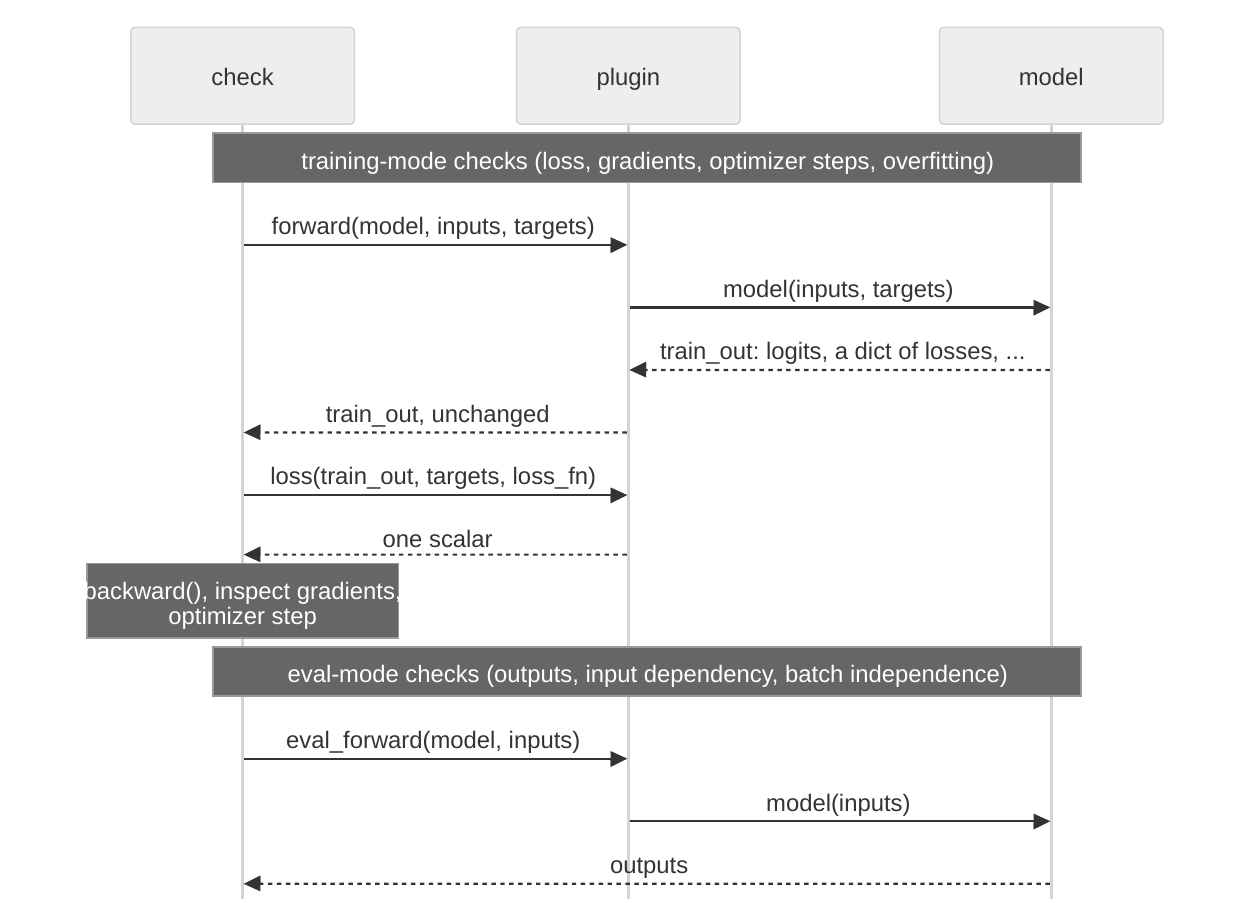}
\caption{How the checks call a plugin. Training-mode checks pass the return value of
\texttt{forward} unchanged to \texttt{loss}; eval-mode checks call \texttt{eval\_forward}. Without
a plugin, both forward methods call \texttt{model(inputs)} and the loss is the user's
\texttt{loss\_fn}.}
\label{fig:pluginflow}
\end{figure}

Some model families are not trained through \texttt{model(inputs)} and a loss function. A
\emph{plugin} bundles how a family runs: \texttt{forward} (training mode, may take the targets),
\texttt{eval\_forward}, \texttt{loss} (which receives whatever \texttt{forward} returned, e.g.\ a
dict of losses, and reduces it to one scalar; Figure~\ref{fig:pluginflow}), the batch dimensions of inputs and outputs (they differ,
e.g., for a sequence-first recurrent classifier), and optionally an optimizer.
\tool{} ships plugins for torchvision detection models (training mode returns a dict of losses),
Hugging Face models (labels are passed to the model, which computes the loss), and common
PyTorch Lightning patterns (\texttt{training\_step} and the first optimizer of
\texttt{configure\_optimizers}, without reproducing the Trainer loop: no strategies, precision
plugins, manual optimization, schedulers or callbacks). Third-party plugins register under a Python entry-point group and are selected by
name; only the requested entry point is imported, so a broken unrelated plugin cannot break a
test suite. A plugin changes how the model is run, never what is checked, so reports stay
comparable across model families.

\section{Evaluation methodology}
\label{sec:method}

We ask:
\RQ{1} How often does \tool{} raise false alarms on healthy models, and what do its findings
on them mean?
\RQ{2} Which silent bugs does it detect, compared with existing practice?
\RQ{3} Do the results generalise beyond popular architectures?
\RQ{4} What does it cost?
\RQ{5} Is it robust to unusual but legitimate setups, deterministic, and does it restore the
state it touches?
\RQ{6} Do false-alarm and detection rates hold on architectures that played no part in
developing and tuning \tool?
\RQ{7} Which real silent bugs are within its reach?
\RQ{8} Does it behave differently on pretrained weights?

\paragraph{Model zoo.} Table~\ref{tab:zoo} summarises the \nModels{} models: classification,
segmentation, detection and video models from torchvision \citep{torchvision2016}, image
classifiers from timm \citep{wightman2019timm}, encoders, decoders (including a mixture of
experts), encoder--decoders, vision, audio and a contrastive model from Hugging Face transformers
\citep{wolf2020transformers}, and hand-written recurrent and attention models. All are randomly
initialised; configurations are reduced (small images, two layers for transformers) so that the
full zoo fits an 8\,GB GPU. Each model comes with a deterministic batch, a loss and the arguments
a user would pass (e.g.\ the detection or Hugging Face plugin). The optimizer is Adam
($\mathrm{lr}=10^{-3}$) and \chk{overfit\_batch} is limited to 100 steps, the budget of the
overfitting baseline; the library's default is 200 steps (Section~\ref{sec:cost} reports both).

\begin{table}[htbp]
\centering\small
\caption{The model zoo (models that pass a smoke test with our configuration).}
\label{tab:zoo}
\begin{tabular}{llrr}
\toprule
Family & Tasks & Models & Parameters \\
\midrule
pytorch & classification, sequence & 4 & 2.9k--58.6k \\
torchvision & classification, detection, segmentation, video & 36 & 352.0k--128.8M \\
timm & classification & 32 & 1.1M--29.4M \\
transformers & audio, classification, contrastive, lm, seq2seq, text-classification & 31 & 19.7k--4.7M \\
\midrule
total & & 103 & \\
\bottomrule
\end{tabular}
\end{table}

\paragraph{Mutation operators.} Table~\ref{tab:mutations} lists the 17 operators and the
equivalent-mutant control. They are implemented generically with PyTorch hooks, so that the same
operator applies to any model: activations of the middle leaf module (in forward order) are
detached, zeroed, mixed across the batch, dropped out in eval mode or given NaN gradients with a
finite forward pass; inputs are partially blinded or attention masks ignored; the optimizer
misses the last layer or has a zero learning rate; a layer declared frozen is trained; the loss
is re-created as a leaf tensor or not reduced; initialisations are scaled. An operator is
\emph{not applicable} when the model lacks the structure it needs (e.g.\ BatchNorm).

\begin{table}[htbp]
\centering\small
\caption{Mutation operators (silent bugs) and the checks designed to catch them.}
\label{tab:mutations}
\begin{tabular}{lp{5.6cm}p{4.2cm}}
\toprule
Operator & Injected bug & Checks designed to catch it \\
\midrule
\texttt{noop} & identity hook (equivalent mutant, control) & none (equivalent mutant) \\
\texttt{unused\_layer} & a layer is registered but never called & \texttt{gradient\_flow}, \texttt{params\_change} \\
\texttt{detach} & the graph is cut (.detach()) in the middle of the network & \texttt{gradient\_flow} \\
\texttt{dead\_layer} & an intermediate activation is replaced by zeros & \texttt{gradient\_flow}, \texttt{input\_dependency}, \texttt{output\_varies} \\
\texttt{batch\_mix} & samples are mixed along the batch dimension mid-network & \texttt{batch\_independence} \\
\texttt{softmax\_batch} & the output is normalised over the batch dimension & \texttt{batch\_independence} \\
\texttt{bn\_batch\_stats} & BatchNorm uses batch statistics at inference & \texttt{batch\_independence} \\
\texttt{input\_ignored} & half of an input's features never reach the model & \texttt{input\_dependency} \\
\texttt{mask\_ignored} & the attention mask is ignored & \texttt{input\_dependency} \\
\texttt{nan\_grad} & a finite forward pass with NaN gradients & \texttt{gradient\_magnitude} \\
\texttt{dropout\_in\_eval} & dropout stays active in eval mode & \texttt{eval\_determinism} \\
\texttt{explode\_init} & weights initialised 10x too large & \texttt{gradient\_magnitude}, \texttt{initial\_loss}, \texttt{output\_finite}, \texttt{overfit\_batch} \\
\texttt{overconfident\_init} & the last layer is initialised 30x too large & \texttt{initial\_loss} \\
\texttt{optimizer\_missing} & the optimizer lacks the last layer's parameters & \texttt{optimizer\_coverage}, \texttt{params\_change} \\
\texttt{lr\_zero} & the learning rate is zero & \texttt{params\_change}, \texttt{overfit\_batch} \\
\texttt{frozen\_breach} & a layer declared frozen is still trained & \texttt{frozen\_params} \\
\texttt{loss\_rewrapped} & the loss is re-created as a new leaf tensor & \texttt{loss}, \texttt{gradient\_flow}, \texttt{params\_change} \\
\texttt{loss\_unreduced} & the loss is not reduced to a scalar & \texttt{loss} \\
\bottomrule
\end{tabular}
\end{table}

\paragraph{Baselines.} \emph{Smoke test}: one training step must run and yield a finite loss
(as Lightning's \texttt{fast\_dev\_run}). \emph{torchtest (2019)}: a faithful reimplementation of
the original library's suite---all trainable variables change after one step, then the outputs
contain no NaN and no inf. \emph{Overfit one batch}: the loss must drop by at least 10\% within
the same 100-step budget. \emph{Training monitor}: the implementations of AUTOTRAINER and DeepDiagnosis target Keras, so we re-implemented
the symptoms they monitor---non-finite loss, gradients or weights, exploding and vanishing
gradients, dying ReLUs, weights that never change, a loss that does not decrease---over the same
100 steps; the thresholds are ours, as the published rules are tied to their implementations.
We did not use TrainCheck \citep{jiang2025traincheck} as a baseline: it needs instrumented traces
of known-good training pipelines to infer its invariants, whereas the question here is what can be
learned from a model and one batch before such a pipeline exists; its published silent errors
serve instead as a source of real bugs (below).

\paragraph{Counting detections.} A healthy model may already receive warnings (and, rarely, a
\texttt{FAIL}), and a baseline may already fail on it; an alarm that the unmodified model also
raises says nothing about the injected bug. Every mutant is therefore compared with its
\emph{healthy counterpart}---the same model, seed and tool version without the mutation. Our
primary metric is \emph{mutation-specific} detection: for \tool{}, strict when some check is
\texttt{FAIL} on the mutant but not on the healthy model, lenient when some check is more severe
on the mutant ($\texttt{PASS}=\texttt{SKIP}<\texttt{WARN}<\texttt{FAIL}$); for a baseline, when
it fails on the mutant and passes on the healthy model. The two definitions aggregate differently:
\tool{} is credited for a new \texttt{FAIL} in \emph{one check} even when its healthy report
already had a \texttt{FAIL} in another, whereas a baseline has one verdict per run. Tools are
therefore compared on a third, like-for-like measure, \emph{report-level strict} detection: the
mutant's report contains a \texttt{FAIL} and the healthy model's report contains none, which is
exactly the rule applied to the baselines. The check-level rates measure, in addition, the new
diagnostic information that a per-check report provides. For comparison with earlier work we also
give the raw rates: any \texttt{FAIL}, and a \texttt{FAIL} or \texttt{WARN} of a check designed for
the bug (Table~\ref{tab:mutations}). The \texttt{noop} operator, an equivalent mutant, checks the
pairing: its mutation-specific detection rate should be zero.

\paragraph{Statistics.} Mutants are nested in models (and models in families), so they are not
independent trials. Rates over all operators come with 95\% percentile intervals from a cluster
bootstrap over models \citep{efron1993bootstrap}: models are resampled with replacement, each
with all of its mutants (2000 replicates). Because all tools run on the same mutants, tools are
compared on paired outcomes: the difference of their mutation-specific rates on the same mutants,
with a cluster-bootstrap interval. We report effect sizes with intervals rather than tests that
would treat individual mutants as independent. Models are in turn nested in architecture families
(several held-out models come from the same timm module, and torchvision's ResNet builder yields
seven zoo models); as a sensitivity analysis, every interval is recomputed with a bootstrap over
families instead of models (timm models grouped by timm module, torchvision models by builder
module, Hugging Face models by model type; \pairHeldFamilies{} families in the held-out zoo).
Per-operator rates, where each model contributes one mutant, carry Wilson intervals
\citep{wilson1927}. The intervals describe the sampling of models, not of operators, which are
fixed by design.

\paragraph{Classifying findings.} We classify every \texttt{FAIL} and every kind of \texttt{WARN}
on a healthy model by hand into three classes: an \emph{incorrect finding} states something that
is false (a false alarm in the strict sense); a \emph{correct but intended} finding states a true
property that the model's authors chose deliberately or that is harmless in context (a contrastive
loss couples the samples of a batch by design); a \emph{genuine defect} is a true finding that
contradicts the intent of the model or the library.
\paragraph{Held-out zoo.} The fixes of Section~\ref{sec:rq1} were prompted by the zoo on which we
then measure false alarms. To measure them on unseen models, we froze \tool{} and selected, with a
fixed random seed and before any run, (i) up to three models from every timm module (architecture
family) that the zoo does not use, with at most 30M parameters, and (ii) up to 40 model types per
task from the Hugging Face auto-model mappings not used in the zoo, in reduced configurations.
Of \hoCandidates{} candidates, \hoModels{} pass the smoke test; the others are incompatible with
the reduced configuration and are listed in the artefacts.

\paragraph{Real bugs.} From the replication package of \citet{shah2024reproducibility} we took
every PyTorch bug that comes with reproduced code (\rbTotal) and classified it as a crash, a
question rather than a bug, a silent bug outside a model tester's scope (data pipeline,
metrics, logging, hyper-parameters, performance) or a silent bug within it. The latter were
ported, with the fix of the accepted answer, into buggy/fixed pairs, keeping model and bug. As a
second independent source we classified, from their descriptions and reproduction scripts, the
\tcTotal{} silent training errors released in the public artefact of TrainCheck
\citep{jiang2025traincheck} (its paper evaluates 20 errors and detects 18 of them; 18 of the 20
are public), and ported those within scope in the same way.

\paragraph{Pretrained weights.} \prN{} zoo architectures were also run with their published
weights and compared with the same architectures randomly initialised, on the healthy model and
all operators.

\paragraph{Realistic bugs.} Generic operators may not look like the bugs people write. We
therefore also wrote \curN{} small programs, each containing one typical silent bug taken from
practitioner reports and our own experience (selected and written by the author of the tool, a
bias that the two independent sources above are meant to offset) (a stale optimizer after replacing a head, the
legacy \texttt{.data}, \texttt{batch\_first} mix-ups, a \texttt{view} instead of a
\texttt{transpose}, an ignored attention mask, a hand-written \texttt{backward} missing a
factor, \ldots), and its fixed version. A tool that flags a fixed version raises a false alarm.

\paragraph{Fuzzing.} To test generalisation beyond popular architectures and to obtain tight
confidence intervals, we generate random architectures (random stacks of linear, residual,
normalisation, dropout, convolution, attention and recurrent blocks on vector or sequence inputs)
and apply every operator to each.

\paragraph{Robustness.} An edge-case study runs \edgeCases{} unusual but legitimate setups (lazy
modules, half precision, \texttt{torch.compile}, tied weights, parametrisations, complex and
integer outputs, packed sequences, closure-based and sparse optimizers, \ldots). Property-based
tests \citep{maciver2019hypothesis} check, on random models, that reports are deterministic and
that nothing observable changes, and a metamorphic relation \citep{segura2016metamorphic}: the
verdicts do not depend on the order of the samples in the batch.

\paragraph{Cost.} Every run records the wall-clock time of each check. In addition, a scaling
experiment measures a full check on GPT-2 models of increasing size against a plain
training step, with and without \chk{overfit\_batch}.

\paragraph{Environment and provenance.} One laptop: NVIDIA RTX 3070 (8\,GB), 16 CPU cores,
PyTorch 2.11, torchvision 0.26, timm 1.0.30, transformers 5.17, Lightning 2.6. PyTorch defaults
(e.g.\ TF32 for convolutions) are kept, so that false alarms are measured under realistic
numerics. The benchmark runs a frozen, non-editable install of \tool{} and every result records
the commit it was produced with. The GPU lane runs torchvision and timm models, a CPU lane the
hand-written and Hugging Face models.

\paragraph{Versions.} \tool{} changed in response to what the evaluation found, so each result
belongs to a version, and a version can only be evaluated without bias on data that played no
part in tuning it. Table~\ref{tab:versions} lists the versions, what each was tuned on and which
results it produced. Only V3 on the held-out zoo is a held-out measurement in the strict sense.

\begin{table}[htbp]
\centering\footnotesize
\caption{Versions of \tool{} in this paper: git commit of the library, with the benchmark snapshots
that recorded it in parentheses where they differ (a snapshot differs from its library commit only
outside the library). \emph{Unbiased on}: data no change of this version was based on.
Table~\ref{tab:provenance} in Appendix~\ref{app:tables} maps every result file to its version.}
\label{tab:versions}
\begin{tabular}{p{0.14\linewidth}p{0.2\linewidth}p{0.33\linewidth}p{0.17\linewidth}}
\toprule
Version & Tuned on & Results & Unbiased on \\
\midrule
V0 \texttt{60e7749} & unit tests, README examples, a first fuzzing run & first zoo run
  (Table~\ref{tab:history}, ``before'') & the zoo \\
V1 \texttt{f5e9636} & V0 + fixes prompted by the zoo and the edge cases & main zoo (healthy,
  mutation, stability), fuzzing, realistic bugs, edge cases, devices, cost & held-out zoo (not
  run) \\
V2 \texttt{d2b1754} (\texttt{1dc4d58}) & V1's zoo results & post-evaluation re-run (Section~\ref{sec:postfix}) & --- \\
V3 \texttt{dac9471} (\texttt{16ebf8a}), frozen & the main zoo & held-out zoo, pretrained weights,
  training monitor, real bugs (Shah et al.) & \textbf{held-out zoo} \\
V4 \texttt{34f102a} (\texttt{441d088}) & the held-out zoo & re-check of both zoos
  (Section~\ref{sec:heldoutfixes}) & a future unseen zoo \\
V5 \texttt{ec61485} (\texttt{be34ab9}) & a first code review (no benchmark data) & re-run of
  both zoos, cost with defaults & a future unseen zoo \\
V6 \texttt{60f93d1} (\texttt{9b259aa}) & a second code review (no benchmark data) & re-run of
  both zoos, cost with defaults, TrainCheck bugs & a future unseen zoo \\
V7 \texttt{ee6d0c7} (\texttt{9028890}) & a third code review (no benchmark data) & re-run of
  both zoos, cost with defaults; the released version & a future unseen zoo \\
\bottomrule
\end{tabular}
\end{table}

\section{Results}
\label{sec:results}

\subsection{RQ1: false alarms on healthy models}
\label{sec:rq1}
With V1, the version of the main zoo evaluation, \nHealthyRuns{} runs on the healthy zoo (\nModels{} models,
three seeds) produced a \texttt{FAIL} in \nHealthyFailRuns{} runs, on \nHealthyFailModels{}
model(s). We inspected every \texttt{FAIL} and every kind of \texttt{WARN}; Table~\ref{tab:findings}
classifies them. Of the \texttt{FAIL}s, one is a genuine defect in a widely used library
(Section~\ref{sec:gcvit}), two are correct but intended findings---parameters that never
receive a gradient with this head, and a batch coupling that a contrastive loss introduces by
design---and one, on timm's EdgeNeXt, is an incorrect finding, fixed after the evaluation
(Section~\ref{sec:postfix}). The warnings are all correct; most describe intended properties
(zero-initialised layers), some questionable defaults (over-confident initialisations).

\begin{table}[htbp]
\centering\small
\caption{Classification of the findings on healthy models (V1).}
\label{tab:findings}
\begin{tabular}{p{0.19\linewidth}lp{0.47\linewidth}}
\toprule
Model(s) & Finding & Explanation \\
\midrule
timm GCViT & \texttt{FAIL} batch indep. & \textbf{Genuine defect}: global queries are tiled across the batch
  (Section~\ref{sec:gcvit}). \\
HF CANINE (classification) & \texttt{FAIL} gradient flow & \textbf{Correct, intended}: the up-sampling
  projection is only used by the sequence-output head; 20 tensors never receive a gradient. Such
  parameters break \texttt{DistributedDataParallel} unless \texttt{find\_unused\_parameters=True}. \\
HF CLIP & \texttt{FAIL} batch indep. & \textbf{Correct, intended}: the contrastive logits compare every
  image with every text of the batch; users skip the check. \\
torchvision ViT, timm ResNet-D, RegNet, LeViT, NF-ResNet, MLP-Mixer, \ldots &
  \texttt{WARN} gradient flow & \textbf{Correct, intended}: zero-initialised heads or residual
  branches give exactly zero gradients at initialisation. \\
torchvision EfficientNet, MNASNet & \texttt{WARN} initial loss & \textbf{Correct, questionable default}: with
  \texttt{num\_classes=10} the classifier initialisation ($\mathcal{U}(\pm 1/\sqrt{C})$ for
  EfficientNet) is over-confident; the initial loss is 2--6$\times$ $\ln C$. \\
torchvision MobileNetV3 & \texttt{WARN} output varies & \textbf{Correct, intended}: in eval mode at
  initialisation, the logits of different inputs differ by $\approx 10^{-10}$ (by $0.04$ in
  training mode); untrained BatchNorm statistics make the network ignore its input. \\
torchvision SqueezeNet, timm Sequencer & \texttt{WARN} input dep. & \textbf{Correct, intended}: border pixels
  are never covered by a stride-2 convolution without padding at this input size. \\
timm EdgeNeXt & \texttt{FAIL} params change & \textbf{Incorrect} (V1):
  a layer scale of $10^{-6}$ makes the gradients of the branch behind it so small that Adam's
  update does not change them in float32; now a \texttt{WARN} about vanishing gradients
  (Section~\ref{sec:postfix}). \\
various & \texttt{WARN} overfit & \textbf{Correct, harmless}: the 100-step budget is too short
  for per-pixel random labels (segmentation) or large detectors. \\
\bottomrule
\end{tabular}
\end{table}

\paragraph{How the evaluation reduced false alarms.} The first run on the zoo, with the version
of \tool{} that had passed its own test suite and a fuzzing study, produced a \texttt{FAIL} in
\histFailRunsBefore{} of \histRunsBefore{} runs on \histFailModelsBefore{} models, and crashed
on one model (Table~\ref{tab:history}). Inspecting them revealed seven problems in \tool{}, all
fixed: (1) the snapshot dropped \texttt{state\_dict.\_metadata}, which MNASNet needs to reload
its weights; (2) a parameter with an exactly-zero gradient was reported as an optimizer failure;
(3) inputs hidden at initialisation by zero-initialised layers were reported as ignored, which
we now re-check through the training objective and after one optimizer step; (4) the same for
detection outputs, which are piecewise constant after non-maximum suppression; (5) the loss at
initialisation was measured in eval mode, where untrained BatchNorm statistics inflate the
logits of deep networks; (6) gradient magnitudes were judged in absolute terms rather than
relative to the parameter; (7) eval-mode input gradients failed for cuDNN RNNs. Real models are
therefore essential to validate a model tester: synthetic architectures (Section~\ref{sec:rq3})
exhibited none of these patterns.

\begin{table}[htbp]
\centering\small
\caption{Healthy zoo, before (V0) and after (V1) the fixes prompted by this evaluation (runs per
status).}
\label{tab:history}
\begin{tabular}{lrrrr}
\toprule
 & \multicolumn{2}{c}{before} & \multicolumn{2}{c}{after} \\
Check & FAIL & WARN & FAIL & WARN \\
\midrule
\texttt{output\_varies} & 0 & 34 & 0 & 4 \\
\texttt{initial\_loss} & 0 & 6 & 0 & 6 \\
\texttt{gradient\_flow} & 3 & 43 & 3 & 44 \\
\texttt{gradient\_magnitude} & 0 & 18 & 0 & 5 \\
\texttt{params\_change} & 49 & 0 & 6 & 44 \\
\texttt{input\_dependency} & 24 & 0 & 0 & 6 \\
\texttt{batch\_independence} & 6 & 0 & 6 & 1 \\
\texttt{overfit\_batch} & 0 & 27 & 0 & 26 \\
\midrule
crashed runs & 3 & & 0 & \\
runs with a FAIL & 63/306 & & 12/308 & \\
models with a FAIL & 21 & & 4 & \\
\bottomrule
\end{tabular}
\end{table}

\subsection{A real bug found in timm}
\label{sec:gcvit}
\chk{batch\_independence} failed on timm's GCViT (timm 1.0.30). Evaluating each image alone
and in a batch of four confirms it: the logits differ by up to $1.0\times10^{-2}$ (mean logit
magnitude $0.21$), against $\le 10^{-7}$ for a ViT. In \texttt{WindowAttentionGlobal}, the
per-image global queries are expanded to the attention windows with
\texttt{q\_global.repeat(B // q\_global.shape[0], 1, 1, 1)}. The windows are laid out image by
image, so \texttt{repeat}---which tiles $q_0, q_1, q_0, q_1, \ldots$---assigns the global query
of one image to windows of another. Replacing it by \texttt{repeat\_interleave} restores batch
independence (difference $3\times10^{-7}$). The bug makes predictions depend on which other
images share the batch, e.g.\ on the evaluation batch size, and it cannot be seen from the loss
or accuracy curves. A reproduction script is part of the artefacts.

\subsection{RQ2: detection of silent bugs in real models}
\label{sec:rq2}
Table~\ref{tab:paired} summarises detection over all operators for the three mutation studies,
counted relative to the healthy counterpart and, for comparison, as raw rates. With the same
aggregation for every tool---a failing report where the healthy model's report does not
fail---\tool{} detects \pairZooTTReportStrict\% of \detZooTTN{} applicable mutants on the zoo
(95\% CI \pairZooTTReportStrictLo--\pairZooTTReportStrictHi), the original torchtest checks
\pairZooClassicPaired\%, the training monitor \pairZooMonitorPaired\%, overfitting one batch
\pairZooOverfitPaired\% and a smoke test \pairZooSmokePaired\%. Counted per check, \tool{}
reports a new \texttt{FAIL} for \pairZooTTPairedStrict\% of the mutants
(\pairZooTTPairedStrictLo--\pairZooTTPairedStrictHi) and a more severe status for
\pairZooTTPairedLenient\% (\pairZooTTPairedLenientLo--\pairZooTTPairedLenientHi); the
difference to the report-level rate is the mutants whose healthy model already failed another
check, where the per-check report still names the new problem.

Pairing barely changes the rates of \tool{}: the raw rates are \detZooTT\% (any \texttt{FAIL})
and \detZooTTlenient\% (a designed check warns or fails), and the \texttt{noop} control drops
from \pairNoopZooTTLenient{} flagged mutants (raw, lenient) to \pairNoopZooTTPairedStrict{}
(strict) and \pairNoopZooTTPairedLenient{} (lenient: two torchvision detectors whose
\chk{overfit\_batch} warning depends on non-deterministic GPU training). The warnings that healthy models already receive are, in other words, not
what detects the injected bugs: the designed checks change status. Pairing matters for the
baselines instead: the original torchtest checks fail on \pairHealthyZooClassicRaw\% of the
healthy models (a parameter that legitimately does not change in one step: behind a
zero-initialised layer, or unused by this head) and the training monitor on \pairHealthyZooMonitorRaw\%, so their raw rates
(\detZooClassic\% and \detZooMonitor\%) overstate what they detect. On the same mutants and
with the same report-level aggregation, \tool{} exceeds the torchtest checks by
\pairDiffReportZooClassic{} percentage points (95\% CI
\pairDiffReportZooClassicLo--\pairDiffReportZooClassicHi) and the monitor by
\pairDiffReportZooMonitor{} (\pairDiffReportZooMonitorLo--\pairDiffReportZooMonitorHi);
Table~\ref{tab:pairedcomparison} in Appendix~\ref{app:tables} gives these and the check-level
differences, which are larger because they also count the per-check information; with a
bootstrap over architecture families the intervals are the same to within one percentage point.

\begin{table}[htbp]
\centering\small
\caption{Detection over all operators (the \texttt{noop} control excluded), relative to the
healthy counterpart and raw, with 95\% intervals from a bootstrap over models; alarm rates on
the healthy models; and the equivalent mutant. The report-level row aggregates \tool{} like the
baselines and is the one to compare across tools; the strict and lenient rows count new findings
per check. Zoo: V1; held-out zoo: V3; random architectures: V1.}
\label{tab:paired}
\resizebox{\linewidth}{!}{%
\begin{tabular}{lrrr}
\toprule
Metric & real zoo (103 models) & held-out zoo (191 models) & random (600 models) \\
\midrule
\multicolumn{4}{l}{\emph{Mutation-specific detection, \% [95\% CI, bootstrap over models]}} \\
torchtest2: FAIL not on the healthy model & 80 [79, 82] & 81 [79, 82] & 81 [80, 81] \\
torchtest2: any severity increase & 91 [89, 92] & 90 [89, 92] & 99 [99, 99] \\
torchtest2: report FAILs, healthy report has no FAIL & 78 [74, 80] & 78 [75, 80] & 81 [80, 81] \\
torchtest (2019): fails, healthy passes & 44 [40, 48] & 40 [37, 43] & 52 [52, 52] \\
training monitor: fails, healthy passes & 43 [38, 48] & 48 [46, 51] & -- \\
overfit one batch: fails, healthy passes & 24 [22, 25] & 22 [22, 23] & 27 [26, 27] \\
smoke test: fails, healthy passes & 3 [2, 3] & 0 [0, 0] & 7 [7, 7] \\
\midrule
\multicolumn{4}{l}{\emph{Alarm on the healthy model, \% of models}} \\
torchtest2 FAIL & 4 (4/103) & 6 (11/191) & 0 (0/600) \\
torchtest2 FAIL/WARN & 30 (31/103) & 23 (44/191) & 26 (157/600) \\
torchtest (2019) & 18 (18/101) & 24 (45/191) & 0 (0/600) \\
training monitor & 23 (24/103) & 13 (24/191) & -- \\
overfit one batch & 2 (2/100) & 0 (0/191) & 0 (0/600) \\
smoke test & 0 (0/101) & 0 (0/191) & 0 (0/600) \\
\midrule
\multicolumn{4}{l}{\emph{\texttt{noop} control (equivalent mutant), \% flagged}} \\
torchtest2: FAIL not on the healthy model & 0 (0/101) & 0 (0/190) & 0 (0/556) \\
torchtest2: any severity increase & 2 (2/101) & 0 (0/190) & 0 (0/556) \\
torchtest2: report FAILs, healthy report has no FAIL & 0 (0/101) & 0 (0/190) & 0 (0/556) \\
torchtest2: any FAIL (raw) & 4 (4/101) & 6 (11/190) & 0 (0/556) \\
torchtest2: FAIL/WARN of a designed check (raw) & 32 (32/101) & 23 (44/190) & 21 (119/556) \\
\midrule
\multicolumn{4}{l}{\emph{Raw detection (any alarm on the mutant), \% [95\% CI] -- for comparability}} \\
torchtest2: any FAIL (raw) & 81 [80, 83] & 83 [82, 84] & 81 [80, 81] \\
torchtest2: FAIL/WARN of a designed check (raw) & 90 [88, 92] & 91 [90, 92] & 99 [99, 99] \\
torchtest (2019): fails (raw) & 61 [58, 65] & 64 [61, 67] & 52 [52, 52] \\
training monitor: fails (raw) & 66 [62, 69] & 61 [59, 63] & -- \\
overfit one batch: fails (raw) & 25 [23, 28] & 22 [22, 23] & 27 [26, 27] \\
smoke test: fails (raw) & 3 [2, 3] & 0 [0, 0] & 7 [7, 7] \\
\bottomrule
\end{tabular}%
}
\end{table}

Table~\ref{tab:detzoo} reports mutation-specific detection per operator on the zoo (the raw
rates are in Table~\ref{tab:detzooraw} and Figure~\ref{fig:detzoo} in Appendix~\ref{app:tables}).
Figure~\ref{fig:checkszoo} in Appendix~\ref{app:tables} shows which checks fire for which
bug. The mutation benchmark itself led to one improvement: mixing the queries of the last
attention layer of DistilBERT changed the logits of other samples by only $2.5\times10^{-6}$ at
initialisation (attention is nearly uniform), below the comparison tolerance. Since untouched
samples normally receive bit-identical outputs, \chk{batch\_independence} now warns about any
change above $10^{-6}$ of the output scale; this raised no warning on healthy models on either
device. Batch mixing, ignored inputs and attention masks, dropout at inference, over-confident
initialisation and training a layer declared frozen are, in effect, caught only by \tool: the
baselines' mutation-specific rates for them are close to zero (Table~\ref{tab:detzoo}); their raw
alarms on these mutants are the alarms they already raise on the unmodified models. None of them
prevents the loss from decreasing, which is why the most common advice---overfit a single
batch---misses them.

\begin{table}[htbp]
\centering\small
\caption{Mutation-specific detection (\%) per operator on the model zoo (V1). \emph{tt2 report}: the
mutant's report has a \texttt{FAIL} and the healthy model's report has none, the aggregation used
for every baseline (a failure on the mutant where the healthy model passes); \emph{tt2 strict}, a
new \texttt{FAIL} in some check; \emph{tt2 lenient}, any increase in severity in some check.
$n$: models the operator applies to.}
\label{tab:detzoo}
\resizebox{\linewidth}{!}{%
\begin{tabular}{lrrrrrrrr}
\toprule
Bug (mutation) & $n$ & tt2 report & tt2 strict & tt2 lenient & torchtest (2019) & monitor & overfit & smoke \\
\midrule
noop (control) & 101 & 0 & 0 & 2 & 0 & 0 & 0 & 0 \\
unused\_layer & 103 & 96 & 99 & 99 & 82 & 77 & 0 & 0 \\
detach & 101 & 96 & 99 & 99 & 82 & 76 & 0 & 0 \\
dead\_layer & 101 & 96 & 99 & 99 & 82 & 76 & 7 & 1 \\
batch\_mix & 94 & 65 & 66 & 83 & 0 & 1 & 0 & 0 \\
softmax\_batch & 93 & 81 & 83 & 100 & 2 & 2 & 2 & 2 \\
bn\_batch\_stats & 42 & 100 & 100 & 100 & 0 & 0 & 0 & 0 \\
input\_ignored & 75 & 96 & 100 & 100 & 5 & 11 & 8 & 8 \\
mask\_ignored & 27 & 93 & 96 & 96 & 0 & 0 & 0 & 0 \\
nan\_grad & 101 & 96 & 100 & 100 & 78 & 76 & 98 & 0 \\
dropout\_in\_eval & 101 & 3 & 3 & 73 & 0 & 2 & 0 & 0 \\
explode\_init & 103 & 10 & 10 & 34 & 4 & 18 & 0 & 0 \\
overconfident\_init & 29 & 0 & 0 & 76 & 0 & 41 & 0 & 0 \\
optimizer\_missing & 103 & 96 & 100 & 100 & 82 & 77 & 5 & 0 \\
lr\_zero & 103 & 96 & 99 & 100 & 82 & 77 & 89 & 0 \\
frozen\_breach & 103 & 89 & 93 & 93 & 0 & 1 & 0 & 0 \\
loss\_rewrapped & 103 & 96 & 100 & 100 & 82 & 77 & 89 & 0 \\
loss\_unreduced & 29 & 100 & 100 & 100 & 93 & 72 & 100 & 100 \\
\midrule
all bugs & & 78 & 80 & 91 & 44 & 43 & 24 & 3 \\
\bottomrule
\end{tabular}%
}
\end{table}

\paragraph{Realistic bugs.} Mutation operators are generic by construction. Table~\ref{tab:curated}
complements them with \curN{} hand-written programs that contain a typical silent bug as people
write it---each paired with its fixed version. \tool{} flags \curTTlenient{} of them
(\curTT{} with a \texttt{FAIL}), the original torchtest checks \curClassic, overfitting one batch
\curOverfit{} and a smoke test \curSmoke; no tool flags any of the fixed versions. One case
illustrates why checking gradients at every step matters: a learnable noise variance becomes
NaN after a few updates, but \texttt{torch.where} hides it from the forward pass, so the loss
stays finite and training silently continues without the noise.

\begin{table}[htbp]
\centering\small
\caption{Hand-written realistic bugs (\ding{51}: detected; $^\dagger$: the fixed version is also
flagged, i.e.\ a false alarm).}
\label{tab:curated}
\resizebox{\linewidth}{!}{%
\begin{tabular}{lcccccc}
\toprule
 & \multicolumn{2}{c}{torchtest2} & & & & \\
Bug & (FAIL) & (FAIL/WARN) & torchtest (2019) & overfit one batch & smoke test & training monitor \\
\midrule
head replaced after optimizer & \ding{51} & \ding{51} & \ding{51} & -- & -- & \ding{51} \\
legacy data detach & \ding{51} & \ding{51} & \ding{51} & -- & -- & \ding{51} \\
mha batch first & \ding{51} & \ding{51} & -- & -- & -- & -- \\
lstm batch first & \ding{51} & \ding{51} & -- & -- & -- & -- \\
view instead of transpose & \ding{51} & \ding{51} & -- & -- & -- & -- \\
softmax over batch & \ding{51} & \ding{51} & -- & -- & -- & -- \\
two tower slice & \ding{51} & \ding{51} & -- & -- & -- & -- \\
attention mask ignored & \ding{51} & \ding{51} & -- & -- & -- & -- \\
frozen backbone trained & \ding{51} & \ding{51} & -- & -- & -- & -- \\
unused layernorm & \ding{51} & \ding{51} & \ding{51} & -- & -- & \ding{51} \\
pairwise distance nan & \ding{51} & \ding{51} & \ding{51} & \ding{51} & -- & \ding{51} \\
noise variance nan & \ding{51} & \ding{51} & -- & -- & -- & \ding{51} \\
dropout at inference & -- & \ding{51} & -- & -- & -- & -- \\
overconfident init & -- & \ding{51} & -- & -- & -- & -- \\
loss not reduced & \ding{51} & \ding{51} & \ding{51} & \ding{51} & \ding{51} & \ding{51} \\
loss rewrapped & \ding{51} & \ding{51} & \ding{51} & \ding{51} & -- & \ding{51} \\
learning rate too small & \ding{51} & \ding{51} & -- & \ding{51} & -- & \ding{51} \\
input overwritten & \ding{51} & \ding{51} & \ding{51} & \ding{51} & -- & \ding{51} \\
optimizer of other instance & \ding{51} & \ding{51} & \ding{51} & \ding{51} & -- & \ding{51} \\
custom backward wrong & \ding{51} & \ding{51} & -- & -- & -- & -- \\
\midrule
detected & 18/20 & 20/20 & 8/20 & 6/20 & 1/20 & 10/20 \\
false alarm on the fixed version & 0/20 &  & 0/20 & 0/20 & 0/20 & 0/20 \\
\bottomrule
\end{tabular}%
}
\end{table}

\subsection{Post-evaluation fixes}
\label{sec:postfix}
The zoo results of V1 exposed three more shortcomings, which we fixed after the evaluation had
finished, giving V2 (library commit \texttt{d2b1754}, benchmark snapshot \texttt{\pfCommit}): (1) the EdgeNeXt false alarm above: a parameter
that does not change because its gradient is vanishingly small relative to its magnitude is
now reported as a warning, like one whose gradient is exactly zero; (2) \texttt{frozen=} names
were matched only against the canonical name of each parameter, so a frozen name that is an
alias of a tied parameter (T5 and BART share their embedding matrix between \texttt{shared} and
the encoder/decoder) was silently ignored---\tool{} now resolves aliases and warns about names
that match no parameter; (3) models whose eval-mode outputs are constant at initialisation (e.g.\
the untrained MobileNetV3 of Table~\ref{tab:findings}) hid a softmax over the batch dimension,
because all outputs stay equal whatever the other samples are---\chk{batch\_independence} now
repeats the comparison after one optimizer step in that case. We re-ran \tool{} alone (the
baselines are unaffected) on all models and operators with V2.
Table~\ref{tab:postfix} in Appendix~\ref{app:tables} compares the paired runs: over \pfN{} applicable mutants the strict
detection rate moves from \pfStrictBefore\% to \pfStrictAfter\% (\pfLenientBefore\% to
\pfLenientAfter\% with warnings), and the number of healthy models with a \texttt{FAIL} from
\pfHealthyFailBefore{} to \pfHealthyFailAfter{} of \pfHealthyN. None of the curated programs (buggy or
fixed) and none of the edge cases changed verdict. The other tables of the main zoo report V1.

\subsection{RQ3: generalisation to random architectures}
\label{sec:rq3}
On \nMutFuzzModels{} random architectures, \tool{} reported \texttt{FAIL} for \detFuzzTT\% of
\detFuzzTTN{} applicable mutants (\detFuzzTTlenient\% with warnings), the original torchtest
checks for \detFuzzClassic\%, overfitting a single batch for \detFuzzOverfit\% and the smoke test
for \detFuzzSmoke\% (Figure~\ref{fig:fuzz}, Table~\ref{tab:detfuzz} in Appendix~\ref{app:tables}). No healthy random
architecture received a \texttt{FAIL} (\faFuzzTT), and neither did the equivalent mutant.
Relative to the healthy counterparts the rates are the same (\pairFuzzTTPairedStrict\% strict,
\pairFuzzTTPairedLenient\% lenient; Table~\ref{tab:paired}). The strict rate is below 100\% only
because three operators are, by design, reported as warnings: dropout at inference,
over-confident and exploding initialisation.

\begin{figure}[htbp]
\centering
\includegraphics[width=\linewidth]{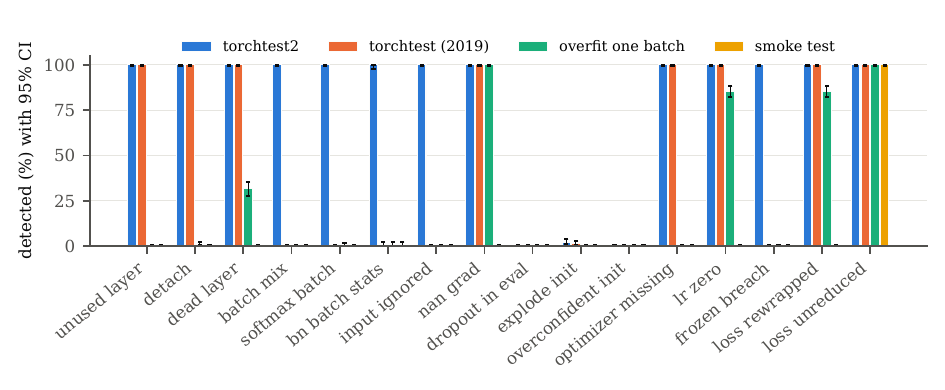}
\caption{Detection rates on random architectures with Wilson 95\% intervals.}
\label{fig:fuzz}
\end{figure}

\subsection{RQ4: cost}
\label{sec:cost}
\begin{figure}[htbp]
\centering
\includegraphics[width=\linewidth]{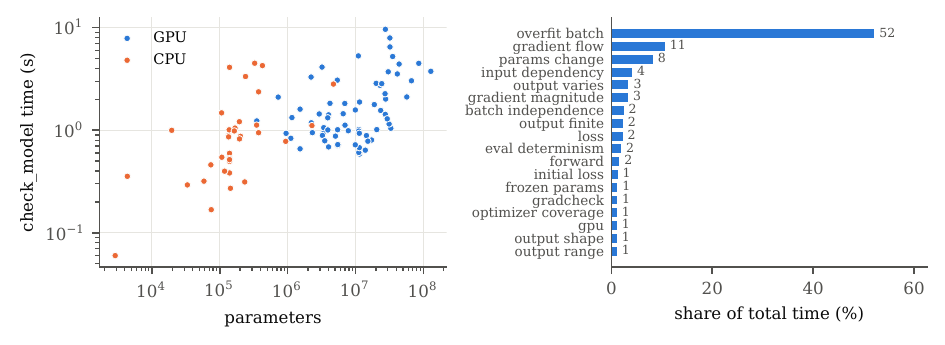}
\caption{Left: time of \texttt{check\_model} vs.\ number of parameters (healthy zoo, seed 0).
Right: share of the total time per check.}
\label{fig:cost}
\end{figure}

\begin{figure}[htbp]
\centering
\includegraphics[width=\linewidth]{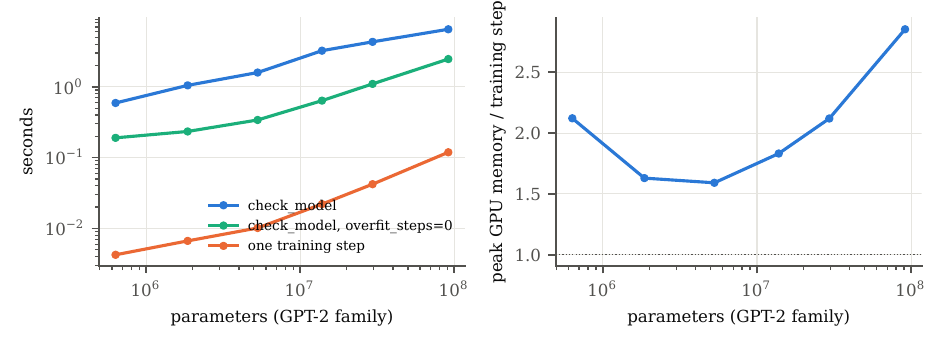}
\caption{Scaling on GPT-2 models: time (left) and peak GPU memory relative to a training step
(right).}
\label{fig:scaling}
\end{figure}

Figure~\ref{fig:cost} shows the time of a complete \chk{check\_model} call on the zoo with the
benchmark configuration (\chk{overfit\_batch} limited to 100 steps). The median is
\costMedian\,s (maximum \costMax\,s), about \costRatio{} times the first training step;
\chk{overfit\_batch} accounts for \costOverfitShare\% of the time and can be shortened or skipped.
With the library's default of 200 steps (V7, the released version, \costSevDefaultModels{} zoo
models), the median is \costSevDefaultMedian\,s (maximum \costSevDefaultMax\,s), against
\costSevDefaultBenchMedian\,s for the same version with 100 steps: \chk{overfit\_batch} stops as
soon as it reaches its target, so the larger budget only costs time on models that need it, and
\costSevDefaultOverfitWarn{} models instead of \costSevDefaultBenchOverfitWarn{} still get a
warning that the target was not reached.
Figure~\ref{fig:scaling} extends this to GPT-2 models of up to \scaleMaxParams{} parameters: the
full check takes \scaleMaxSeconds\,s (\scaleMaxSecondsNoOverfit\,s without overfitting) where a
training step takes \scaleMaxStep\,s; peak GPU memory stays within \scaleMemRatio$\times$ that of
a training step, because the snapshot is kept in host memory. Larger models (210M parameters
and more) did not fit the 8\,GB GPU of our machine.

\subsection{RQ5: robustness, determinism and restored state}
\label{sec:fixes}
The edge-case study found \edgeIssuesBefore{} problems in \tool{} before this evaluation
(Table~\ref{tab:edge} in Appendix~\ref{app:tables}): a crash on lazy modules; false alarms on complex-valued outputs, on
zero-padded sequences, on models that validate their inputs (sorted lengths), and on models that
modify their inputs in place; failures with closure-based (L-BFGS) and sparse optimizers;
duplicated reports when the training forward pass fails; a loss re-created as a new leaf tensor
that passed the \chk{loss} check; and Hugging Face models configured to return tuples. All were
fixed; after the fixes all \edgeCases{} cases pass. Pure float16 training with Adam is correctly reported as diverging
(Adam's $\epsilon=10^{-8}$ underflows in float16); the message now explains why.
No run left a side effect visible to the benchmark's fingerprint (\nSideEffectRuns{} runs with a
difference in parameters, buffers, gradients, flags, optimizer state, PyTorch RNG state or input
tensors); the gaps it could not see are discussed in Section~\ref{sec:execution}. Repeated runs
gave identical
reports (\nDetIdentical/\nDetPairs); property-based tests on 400 random models found no
counterexample to determinism, the restoration of the PyTorch state or batch-order invariance. Across five seeds,
\stabSameVerdict{} of \stabGroups{} (model, operator) pairs received the same \texttt{FAIL}
verdict, and \stabIdentical{} identical reports. Differential testing across devices ran
\devPairs{} setups (25 models, healthy and five operators) on both the CPU and the GPU:
\devFailDiff{} differed in a \texttt{FAIL} verdict and \devWarnDiff{} only in a warning near a
heuristic threshold---e.g.\ the eval-mode outputs of an untrained EfficientNet are exactly
constant on the CPU, but not under the GPU's TF32 arithmetic.

\subsection{RQ6: unseen architectures}
\label{sec:heldout}
Table~\ref{tab:heldout} compares the frozen version on the zoo it was tuned on and on the
\hoModels{} unseen models. On the latter, \hoHeldFail{} models (\hoHeldFailPct\%) received a
\texttt{FAIL} with seed 0 (\hoAllSeedsBefore{} with any of three seeds), against \hoTunedFail{}
on the tuned zoo. We classified each of them (Table~\ref{tab:heldoutfindings}): eight are
correct findings, among them three Hugging Face models whose outputs do not depend on the
attention mask at all (YOSO, MRA, CPM-Ant, besides FNet, which has no attention by design), so
that padding tokens change their predictions; five are false alarms and one is an artefact of
our reduced configuration. Detection generalises (Table~\ref{tab:paired}): with the same
report-level aggregation for every tool, \tool{} detects \pairHeldTTReportStrict\% of
\detHeldTTN{} mutants (95\% CI \pairHeldTTReportStrictLo--\pairHeldTTReportStrictHi), the
original torchtest checks \pairHeldClassicPaired\%, the training monitor \pairHeldMonitorPaired\%,
overfitting one batch \pairHeldOverfitPaired\% and the smoke test \pairHeldSmokePaired\%. Per
check, \tool{} reports a new \texttt{FAIL} for \pairHeldTTPairedStrict\%
(\pairHeldTTPairedStrictLo--\pairHeldTTPairedStrictHi) and a more severe status for
\pairHeldTTPairedLenient\% (\pairHeldTTPairedLenientLo--\pairHeldTTPairedLenientHi). No
\texttt{noop} mutant is detected by \tool{} under any of these definitions
(\pairNoopHeldTTPairedLenient; per operator in
Table~\ref{tab:detheldoutpaired} in Appendix~\ref{app:tables}). The raw rates are
\detHeldTT\% (any \texttt{FAIL}) and \detHeldTTlenient\% (a designed check warns) for \tool{},
\detHeldClassic\% and \detHeldMonitor\% for the torchtest checks and the monitor
(Table~\ref{tab:detheldout} in Appendix~\ref{app:tables}); the baselines' raw rates are inflated
by their alarms on the healthy models: \faHeldClassic{} for the torchtest checks and
\faHeldMonitor{} for the monitor, against \faHeldTT{} for \tool. With report-level aggregation
for every tool, the advantage of \tool{} over the torchtest checks is
\pairDiffReportHeldClassic{} percentage points (95\% CI
\pairDiffReportHeldClassicLo--\pairDiffReportHeldClassicHi) and over the monitor
\pairDiffReportHeldMonitor{} (\pairDiffReportHeldMonitorLo--\pairDiffReportHeldMonitorHi); the
check-level differences, which also credit the per-check information, are
\pairDiffStrictHeldClassic{} and \pairDiffStrictHeldMonitor{} points. Several held-out models
share an architecture family, so these intervals could understate the dependence between
mutants; with a bootstrap over the \pairHeldFamilies{} families instead of the models the
report-level differences have the intervals
\pairDiffReportHeldClassicFamLo--\pairDiffReportHeldClassicFamHi{} and
\pairDiffReportHeldMonitorFamLo--\pairDiffReportHeldMonitorFamHi, and the report-level rate of
\tool{} \pairHeldTTReportStrictFamLo--\pairHeldTTReportStrictFamHi.

\begin{table}[htbp]
\centering\small
\caption{Healthy models with a \texttt{FAIL} or \texttt{WARN} (seed 0): the zoo \tool{} was tuned
on, the unseen held-out zoo, and the held-out zoo after the fixes of
Section~\ref{sec:heldoutfixes}.}
\label{tab:heldout}
\begin{tabular}{lrrr}
\toprule
Zoo & models & with a FAIL (95\% CI) & with a WARN \\
\midrule
tuned zoo & 103 & 3 (3\%, 1--8) & 19 (18\%) \\
held-out zoo & 191 & 11 (6\%, 3--10) & 35 (18\%) \\
held-out zoo, after the fixes & 191 & 10 (5\%, 3--9) & 36 (19\%) \\
\bottomrule
\end{tabular}
\end{table}

\begin{table}[htbp]
\centering\small
\caption{Classification of the held-out models with a \texttt{FAIL} in any seed.}
\label{tab:heldoutfindings}
\begin{tabular}{p{0.22\linewidth}p{0.16\linewidth}p{0.47\linewidth}}
\toprule
Model(s) & Finding & Explanation \\
\midrule
HF DINOv2 (2 variants), MarkupLM, LED & \texttt{FAIL} gradient flow & \textbf{True}: parameters
  used only for other inputs (the mask token of masked image modelling, XPath embeddings,
  global-attention projections without a global attention mask). \\
HF MRA & \texttt{FAIL} gradient flow, input dep. & \textbf{True}: no gradient reaches the query
  and key projections, and the attention mask has no effect. \\
HF YOSO, CPM-Ant, FNet & \texttt{FAIL} input dep. & \textbf{True}: the outputs do not depend on
  \texttt{attention\_mask} (also with the default widths); for FNet this is by design. \\
timm LowFormer & \texttt{FAIL} batch indep. & \textbf{False alarm}: TF32 convolutions round a
  sample depending on its neighbours ($1.7\times10^{-4}$ of the output scale; $0$ without TF32).
  Fixed. \\
HF M2M-100, SeamlessM4T (2 versions) & \texttt{FAIL} gradient flow & \textbf{False alarm}:
  LayerDrop skipped the layer in this pass. Fixed. \\
HF DINOv2 & \texttt{FAIL} input dep. & \textbf{False alarm}: $14\times14$ patches leave an
  8-pixel border of a $64\times64$ image uncovered; the border rule allowed 2 pixels. Fixed. \\
HF MobileBERT & \texttt{FAIL} input dep. & \textbf{False alarm}: the attention mask changes the
  outputs by only $10^{-6}$ at initialisation. Not fixed. \\
HF RWKV & \texttt{FAIL} forward & \textbf{Artefact}: our reduced configuration breaks the
  eval-mode forward pass. \\
\bottomrule
\end{tabular}
\end{table}

\paragraph{Fixes after the held-out run.}
\label{sec:heldoutfixes}
We then fixed the three causes of the false alarms that are not specific to one model: a
coupling seen under TF32 is re-checked without TF32; the uncovered border of a spatial input may
be up to a quarter of the dimension; parameters without a gradient are re-checked with other
random seeds. With the fixed version, \hoHeldFixedFail{} of the \hoHeldFixedN{} unseen healthy
models receive a \texttt{FAIL} with seed 0 (\hoAllSeedsAfter{} with any seed), and on the
original zoo \hoMainFailAfter{} of \hoMainMutants{} mutants are flagged, against
\hoMainFailBefore{} before: detection is unchanged. These numbers are, again, no longer
held-out; a further unseen zoo would be needed to measure the fixed version without bias.

The fixes of the code review (V5, Section~\ref{sec:execution}) were made without looking at
benchmark data; we re-ran V5 on every operator of both zoos. On the main zoo, compared with V4,
\rvMainFailChanged{} of \rvMainCompared{} runs changed their \texttt{FAIL} verdict, both as
intended: two torchvision detectors with dropout at inference return a different number of boxes
in two eval passes, which V4 compared tensor by tensor until the comparison raised an error,
reported as a \texttt{FAIL}; V5 compares the structure first and reports it as the
\chk{eval\_determinism} warning it is. The other changes of status (\rvMainStatusChanged{} in
all) are \chk{overfit\_batch} warnings near the 10\% target on GPU-trained detectors and
classifiers, and an untrained RepVGG whose outputs are constant on the CPU (V4 ran it there) but
not on the GPU. On the held-out zoo the comparison is with V3, so it also shows the effect of V4:
\rvHeldFailChanged{} of \rvHeldCompared{} runs changed their \texttt{FAIL} verdict, all through
V4's fixes---the border band of DINOv2, the TF32 re-check of LowFormer---and the TF32 re-check
also turned three \texttt{batch\_mix} mutants whose coupling is comparable to TF32 rounding
(efficientvit, LowFormer, RDNet) from \texttt{FAIL} into a warning or a pass, the price of
removing that false alarm. Overall, V5 detects \rvHeldReportStrict\% of the held-out mutants at
report level (\rvHeldPairedStrict\% strictly and \rvHeldPairedLenient\% leniently per check),
as V3 did,
and \rvHeldHealthyFail{} healthy held-out models receive a \texttt{FAIL} with seed 0. These are
not held-out measurements of V5.

V6 fixes the edge cases found by a second review (a lazy model whose
optimizer was passed in is now checked together with a copy of that optimizer, dict outputs are
compared by key, a plugin's output batch dimension is kept when the caller sets
\texttt{batch\_dim}); again no benchmark data was looked at. Re-run on every operator of both
zoos, V6 changed \rlMainFailChanged{} of \rlMainCompared{} \texttt{FAIL} verdicts on the main
zoo and \rlHeldFailChanged{} of \rlHeldCompared{} on the held-out zoo compared with V5
(\rlMainStatusChanged{} and \rlHeldStatusChanged{} statuses in all, each a warning near a
threshold: RepVGG's outputs, exactly constant at initialisation on the CPU but not on the GPU
(the re-runs assign models to the CPU or the GPU lane dynamically), \chk{overfit\_batch} near
its target on GPU-trained detectors, and weak couplings of about $10^{-6}$ of the output scale
in EdgeNeXt and LowFormer). Its mutation-specific rates are
\rlMainReportStrict\% at report level (\rlMainPairedStrict\% and \rlMainPairedLenient\% per
check) on the main zoo and \rlHeldReportStrict\% (\rlHeldPairedStrict\% and
\rlHeldPairedLenient\%) on the held-out zoo.

V7, the released version, also restores buffers registered with \texttt{persistent=False},
which \texttt{state\_dict()} leaves out, so a forward pass that updates one (e.g.\ a counter) no
longer leaves it changed. Re-run on both zoos, it changed \rsMainFailChanged{} of
\rsMainCompared{} and \rsHeldFailChanged{} of \rsHeldCompared{} \texttt{FAIL} verdicts compared
with V6 (\rsMainStatusChanged{} and \rsHeldStatusChanged{} statuses, again warnings near a
threshold: \chk{overfit\_batch} near its target on GPU-trained detectors and classifiers, and
LowFormer's couplings of about $10^{-6}$ of the output scale), and detects
\rsMainReportStrict\% (main) and \rsHeldReportStrict\% (held-out) of the mutants at report level.

\subsection{RQ7: real bugs}
Of the \rbTotal{} PyTorch bugs with reproduced code in the dataset of
\citet{shah2024reproducibility}, \rbCrash{} raise an exception---visible without any tool---and
\rbNotBug{} are questions rather than bugs. Of the remaining nine silent bugs, \rbOutScope{} lie
outside the reach of a model tester (a running loss that is never reset, inputs and labels
shuffled independently, an accuracy computed on broadcast shapes, \ldots) and \rbInScope{}
within it. \tool{} flags all \rbInScope{} (\rbTT{} with a \texttt{FAIL}: a log-softmax over a
single output unit, and a mixture model whose means are rebuilt from a detached Python float;
softmax before \texttt{nn.CrossEntropyLoss} only slows training and gets a warning), with no
false alarm on the fixed versions (Table~\ref{tab:realbugs}).

Of the \tcTotal{} silent training errors released in TrainCheck's public artefact
\citep{jiang2025traincheck},
\tcOutScope{} are outside a model tester's reach: data loss in GPU peer-to-peer transfers,
distributed-training and DeepSpeed bugs, compiler bugs, a missing \texttt{zero\_grad} in the
user's training loop, data-loader seeding, generation and performance issues
(Table~\ref{tab:tcscope} in Appendix~\ref{app:tables}). \tcInScope{} are within reach. We ported
\tcPorted{} with their fixes; the third, MMPretrain stages that are frozen only when
\texttt{train()} is called, needs a 2022 version of that library and would be visible to
\chk{frozen\_params} only if the user declares the frozen stages. \tool{} (V6) flags both ported
bugs with warnings of the designed checks and neither fixed version (Table~\ref{tab:tcbugs}): a
FlashAttention-2 module that keeps dropout active in eval mode (\chk{eval\_determinism}), and a
transfer-learning model whose new head is frozen together with the backbone, so that only the
BatchNorm parameters train (\chk{overfit\_batch} does not reach its target; the fixed version
does). None of the baselines flags either bug, and TrainCheck reports both in its own evaluation
\citep{jiang2025traincheck}, from reference traces rather than from a single batch. Both
sources point the same way, and the small numbers are themselves a result: most real silent bugs
are in the code around the model---the training loop, the data pipeline, distributed training or
the framework---where a pre-training model tester cannot see them.

\begin{table}[htbp]
\centering\small
\caption{Real silent bugs within scope, from \citet{shah2024reproducibility} (\ding{51}:
detected; $^\dagger$: the fixed version is also flagged).}
\label{tab:realbugs}
\resizebox{\linewidth}{!}{%
\begin{tabular}{lcccccc}
\toprule
 & \multicolumn{2}{c}{torchtest2} & & & & \\
Bug & (FAIL) & (FAIL/WARN) & torchtest (2019) & overfit one batch & smoke test & training monitor \\
\midrule
SO 71457035 & \ding{51} & \ding{51} & \ding{51} & \ding{51}$^{\dagger}$ & -- & \ding{51}$^{\dagger}$ \\
SO 61781193 & -- & \ding{51} & -- & -- & -- & -- \\
SO 71765132 & \ding{51} & \ding{51} & \ding{51} & \ding{51} & -- & \ding{51} \\
\midrule
detected & 2/3 & 3/3 & 2/3 & 2/3 & 0/3 & 2/3 \\
false alarm on the fixed version & 0/3 &  & 0/3 & 1/3 & 0/3 & 1/3 \\
\bottomrule
\end{tabular}%
}
\end{table}

\subsection{RQ8: pretrained weights}
On the \prN{} pretrained models, no healthy model received a \texttt{FAIL}, and \tool{} flagged
\prDet\% of \prMutants{} mutants, against \prDetRandom\% for the same architectures randomly
initialised. We conjecture that trained weights make outputs vary more with the input and
between samples, which helps the checks that compare outputs; we did not verify this. \chk{initial\_loss} is, as expected, meaningless for pretrained
heads and should be skipped for them.

\section{Discussion}
\paragraph{What the severity model buys.} Treating all findings as failures would raise the
strict detection rate but also the false-alarm rate: dropout at inference and unusual
initialisations can be intended. Warnings are common on healthy models
(\pairHealthyHeldTTLenient\% of the held-out zoo), yet counting them relative to the healthy
counterpart costs \tool{} almost nothing (Table~\ref{tab:paired}): a bug turns the designed check
from \texttt{PASS} into \texttt{WARN} or \texttt{FAIL}, while the pre-existing warnings of the
healthy model are about something else. For users this suggests reading a report as a
difference---a warning that appears after a change is more telling than one that was always
there---and, in a test suite, acknowledging the known warnings of a model with \texttt{skip}.
Our results support keeping uncertain findings as warnings while making \texttt{strict=True}
available.

\paragraph{Unused parameters are not always bugs---but they matter.} The \texttt{FAIL}s on
healthy models concern parameters that really never receive a gradient. Whether that is a bug
depends on intent; it always wastes memory and breaks \texttt{DistributedDataParallel} without
extra configuration. A \texttt{frozen=} declaration or a \texttt{skip} lets users acknowledge them.

\paragraph{Threats to validity.} \emph{Construct}: mutation operators are proxies for real bugs;
we derived them from fault taxonomies and practitioner folklore, but their distribution differs
from real bug distributions, and fault benchmarks themselves are often unrealistic
\citep{jahangirova2024realfaults}. The real bugs of RQ7 address this only partly: few of them are
within scope. \emph{Internal}: the benchmark was written by the authors of the tool; to mitigate
bias the baselines are faithful to their references, the held-out zoo was committed before any
run, all raw results are released, detections are counted relative to the healthy counterpart,
and the equivalent-mutant control measures spurious detections. The realistic bugs were selected
and written by the author of the tool; the two independent sources of real bugs offset this
only partly, because few of their bugs are within scope. The training monitor re-implements
published symptoms with our thresholds, not the original tools, and TrainCheck could not be run
as a baseline because it needs traces of known-good pipelines. \emph{Conclusion}: mutants of one
model are correlated, so all intervals over operators come from a bootstrap over models, and
tools are compared on paired outcomes; the intervals do not cover the choice of operators. \emph{External}: most models are randomly initialised and reduced in size; a
pretrained subset (RQ8) behaves similarly. Results come from a single machine. One mutant is excluded: the
operator that mixes the batch could not be applied to the PyTorch \texttt{nn.Transformer} because the encoder switches to its nested-tensor fast path, a limitation of our
harness rather than of \tool.

\paragraph{Limitations.} \chk{batch\_independence} detects leaks \emph{between} samples, not
data scrambled \emph{within} a sample; vanishing gradients are not yet flagged (a sigmoid CNN
shows a $10^{3}\times$ gradient-norm ratio between first and last layer without a warning;
only parameters whose update is lost to rounding are flagged); the batch checks need a batch
dimension, given per side when inputs and outputs differ, but not per output tensor; and the
restoration guarantee covers PyTorch state, not arbitrary Python state changed by the model's
own code (Section~\ref{sec:execution}).

\section{Conclusion}
\tool{} turns practitioner folklore about silent DL bugs into 18 checks behind one call that
restores the PyTorch state it touches. On real models, random architectures and hand-written
realistic bugs it detects far more injected silent bugs than the checks it replaces---also when
every tool is credited only with alarms that the healthy model does not raise---at the cost of
seconds, and it found a real bug in a widely used model library. Two independent collections of
real silent bugs show the limits of the approach: most such bugs live outside the model. The evaluation improved the tool itself: running it on
\nModels{} real models exposed false alarms that neither its own test suite nor 600 random
architectures had revealed, and a frozen version confirmed its detection rates on unseen
architecture families, where it also exposed models that ignore their attention mask. We believe
that this is the most transferable lesson for tools that test DL code: validate them on the zoo
of models people actually use, and keep part of that zoo unseen.

\paragraph{Reproducibility.} The replication package---the library as evaluated, the
benchmark, all raw results and the scripts that regenerate every number---is archived on Zenodo at
\url{https://doi.org/10.5281/zenodo.23301395}; the library itself is developed at
\url{https://github.com/Hrehory/torchtest2} and installable with \texttt{pip install torchtest2}.
In the package, \texttt{paper/reproduce.sh} regenerates every table, figure and
number of this paper from the released raw results, compares them with the published ones and
builds the paper in minutes without a GPU; with \texttt{rerun-cpu} it also re-runs the buggy/fixed
studies with the library version that produced each of them and compares every verdict, and
with \texttt{rerun-all} it re-runs every study from scratch (about 30 hours on one GPU), each with
its version and on its device, and compares every verdict, table and number. The
shell scripts in \texttt{paper/bench} run the zoo phases: the main zoo (\texttt{pipeline.sh}),
the held-out zoo and its re-checks (\texttt{night2.sh}, \texttt{post2.sh}) and the re-runs of
V5, V6 and V7 (\texttt{night3.sh}, \texttt{night4.sh}, \texttt{night5.sh}). The
command \texttt{python -m bench.analyze} regenerates every table, figure and number in this paper
from \texttt{paper/results}. Every record of the benchmark phases carries the commit that
produced it; a few early studies (fuzzing, the realistic bugs, devices, scaling, the real bugs
of Shah et al.) did not record it, and their version comes from the benchmark's run log, as
marked in Table~\ref{tab:provenance}. The analysis maps each record to its version
(Tables~\ref{tab:versions} and~\ref{tab:provenance}) and rejects a record from an unknown
commit.

\bibliographystyle{plainnat}
\bibliography{refs}

\clearpage
\appendix
\setcounter{table}{0}\renewcommand{\thetable}{A\arabic{table}}
\setcounter{figure}{0}\renewcommand{\thefigure}{A\arabic{figure}}
\section{Additional tables and figures}
\label{app:tables}

This appendix collects the detailed tables and figures that the results refer to.

\begin{table}[htbp]
\centering\scriptsize
\caption{Provenance: every result file used in this paper, by version (number of records in
parentheses). Each record carries the commit of the benchmark snapshot that produced it, except
the files marked $^\ast$, whose scripts did not record it: their version comes from the
benchmark's run log. Generated by \texttt{bench/versions.py}, which rejects a record from a
commit it does not list.}
\label{tab:provenance}
\begin{tabular}{llp{0.62\linewidth}}
\toprule
Version & Library (snapshots) & Result files (records) \\
\midrule
V0 & \texttt{60e7749} (\texttt{aa2940a}) & \texttt{archive/healthy.cpu (245), archive/healthy.gpu (147), archive/healthy.gpu (318), preflight.cpu (35), preflight.gpu (68)} \\
V1 & \texttt{f5e9636} & \texttt{curated$^\ast$ (160), devices$^\ast$ (150), fuzz.part0$^\ast$ (11400), fuzz.part1$^\ast$ (11400), fuzz.part2$^\ast$ (11400), fuzz.part3$^\ast$ (11400), healthy.cpu (244), healthy.gpu (467), mutation.cpu (2520), mutation.cpuhelper (909), mutation.gpu (4032), scaling$^\ast$ (8), stability.cpu (180), stability.gpu (180)} \\
V2 & \texttt{d2b1754} (\texttt{1dc4d58}) & \texttt{postfix.cpu (665), postfix.cpuhelper (570), postfix.gpu (734)} \\
V3 & \texttt{dac9471} (\texttt{16ebf8a}) & \texttt{curated\_v2$^\ast$ (200), heldout/healthy.cpu (576), heldout/healthy.gpu (952), heldout/mutation.cpu (5868), heldout/mutation.gpu (5371), heldout/mutation.gpugpu2 (5400), heldout/mutation.gpuhelp (683), heldout/preflight.cpu (90), heldout/preflight.gpu (119), monitor.cpu (665), monitor.cpuhelper (70), monitor.gpu (840), monitor.gpugpu2 (588), pretrained/healthy.gpu (232), pretrained/mutation.gpu (2610), pretrained/preflight.gpu (29), realbugs$^\ast$ (30), tuning.cpu (665), tuning.cpuhelper (541), tuning.gpu (760)} \\
V4 & \texttt{34f102a} (\texttt{441d088}) & \texttt{heldout/recheck.cpu (216), heldout/recheck.gpu (357), tuning2.cpu (665), tuning2.cpuhelper (285), tuning2.gpu (1007)} \\
V5 & \texttt{ec61485} (\texttt{be34ab9}) & \texttt{defaults.cpu (35), defaults.cpuhelper (25), defaults.gpu (43), heldout/review.cpu (1368), heldout/review.gpu (2261), review.cpu (665), review.cpuhelper (556), review.gpu (741)} \\
V6 & \texttt{60f93d1} (\texttt{9b259aa}) & \texttt{heldout/release.cpu (1368), heldout/release.gpu (2261), release.cpu (665), release.cpuhelper (570), release.gpu (741), release\_defaults.cpu (35), release\_defaults.cpuhelper (25), release\_defaults.gpu (43), traincheck\_bugs (20)} \\
V7 & \texttt{ee6d0c7} (\texttt{9028890}) & \texttt{heldout/release7.cpu (1368), heldout/release7.gpu (2261), release7.cpu (665), release7.cpuhelper (570), release7.gpu (741), release7\_defaults.cpu (35), release7\_defaults.cpuhelper (26), release7\_defaults.gpu (43)} \\
\bottomrule
\end{tabular}
\end{table}

\begin{table}[htbp]
\centering\small
\caption{\tool{} minus each baseline, on the same mutants, both counted relative to the healthy
counterpart: difference of the detection rates (percentage points) with 95\% intervals from a
paired cluster bootstrap over models and, as a sensitivity analysis, over architecture families.
\emph{Report-level}: \tool{} aggregated like the baselines (a failing report where the healthy
report does not fail), the like-for-like comparison; \emph{strict} and \emph{lenient}: a new
\texttt{FAIL} or a more severe status in some check of \tool, which also credits its per-check
information.}
\label{tab:pairedcomparison}
\resizebox{\linewidth}{!}{%
\begin{tabular}{llrllrllrll}
\toprule
 & & \multicolumn{3}{c}{$\Delta$ report-level, pp} & \multicolumn{3}{c}{$\Delta$ strict, pp} & \multicolumn{3}{c}{$\Delta$ lenient, pp} \\
\cmidrule(lr){3-5}\cmidrule(lr){6-8}\cmidrule(lr){9-11}
Study & Baseline & $\Delta$ & models & families & $\Delta$ & models & families & $\Delta$ & models & families \\
\midrule
real zoo & torchtest (2019) & 34 & [29, 38] & [29, 38] & 36 & [33, 40] & [33, 40] & 47 & [44, 51] & [44, 51] \\
real zoo & overfit one batch & 54 & [50, 57] & [50, 58] & 57 & [55, 59] & [55, 59] & 67 & [65, 70] & [65, 70] \\
real zoo & smoke test & 75 & [72, 78] & [71, 78] & 78 & [76, 79] & [76, 79] & 88 & [87, 90] & [87, 90] \\
real zoo & training monitor & 34 & [28, 40] & [28, 40] & 37 & [32, 42] & [32, 42] & 47 & [43, 52] & [42, 53] \\
\midrule
held-out zoo & torchtest (2019) & 37 & [34, 41] & [33, 42] & 41 & [38, 44] & [37, 45] & 50 & [48, 53] & [47, 54] \\
held-out zoo & overfit one batch & 55 & [52, 58] & [52, 58] & 59 & [57, 60] & [57, 60] & 68 & [67, 70] & [66, 70] \\
held-out zoo & smoke test & 78 & [75, 80] & [74, 80] & 81 & [79, 82] & [79, 82] & 90 & [89, 92] & [89, 92] \\
held-out zoo & training monitor & 29 & [26, 33] & [26, 33] & 32 & [30, 35] & [30, 35] & 42 & [39, 45] & [39, 45] \\
\midrule
random & torchtest (2019) & 28 & [28, 28] & -- & 28 & [28, 28] & -- & 47 & [47, 47] & -- \\
random & overfit one batch & 54 & [53, 54] & -- & 54 & [53, 54] & -- & 73 & [72, 73] & -- \\
random & smoke test & 74 & [74, 74] & -- & 74 & [74, 74] & -- & 92 & [92, 93] & -- \\
\bottomrule
\end{tabular}%
}
\end{table}

\begin{table}[htbp]
\centering\small
\caption{The silent training errors released in TrainCheck's public artefact
\citep{jiang2025traincheck},
classified by whether a pre-training model tester can see them.}
\label{tab:tcscope}
\resizebox{\linewidth}{!}{%
\begin{tabular}{llp{0.5\linewidth}}
\toprule
Bug & Scope & Root cause \\
\midrule
baichuan2-86 & out of scope & hardware: GPU P2P data loss \\
deepspeed-1801 & out of scope & framework: BF16 optimizer gradient clipping desynchronises tensor-parallel weights \\
deepspeed-5794 & out of scope & framework: MoE all\_to\_all with inconsistent args \\
lightning-thunder-725 & out of scope & compiler: autocast not applied to non-torch symbols \\
mmpretrain-702 & in scope (not ported) & framework: frozen\_stages applied only in train(), not in \_\_init\_\_ (frozen\_params with frozen=) \\
pytorch-104336 & out of scope & distributed: DDP hooks lost by .cpu()/.cuda() \\
pytorch-115607 & out of scope & compiler: dynamo optimizer step with None gradients \\
pytorch-51800 & out of scope & user evaluation: spectral norm is initialised by the first training forward; evaluating before it misleads the user, the model itself is correct \\
pytorch-84803 & out of scope & hardware: GPU P2P data loss \\
pytorch-96600 & out of scope & hardware: GPU P2P data loss \\
pytorch-forum-84911 & in scope (ported) & user code: every layer but BatchNorm frozen, including the new head (plus an input resize, a data pipeline bug) \\
stackoverflow-60335387 & out of scope & training loop: optimizer.zero\_grad() missing in the user's loop \\
stackoverflow-67180955 & out of scope & data pipeline: data loader workers share the NumPy seed \\
transformers-17877 & out of scope & performance: weights initialised repeatedly (slow start) \\
transformers-23723 & out of scope & generation: max\_new\_tokens computed before prompt truncation \\
transformers-33844 & in scope (ported) & model code: FlashAttention-2 attention keeps dropout active in eval mode \\
transformers-34204 & out of scope & data pipeline: input processor returns only the first batch item \\
x-jxmnop-ddp-out-of-sync & out of scope & distributed: DDP forward invoked on the wrapped module \\
\bottomrule
\end{tabular}%
}
\end{table}

\begin{table}[htbp]
\centering\small
\caption{The TrainCheck errors within scope, ported as buggy/fixed pairs (\tool{} V6;
\ding{51}: detected; $^\dagger$: the fixed version is also flagged).}
\label{tab:tcbugs}
\resizebox{\linewidth}{!}{%
\begin{tabular}{lcccccc}
\toprule
 & \multicolumn{2}{c}{torchtest2} & & & & \\
Bug & (FAIL) & (FAIL/WARN) & torchtest (2019) & overfit one batch & smoke test & training monitor \\
\midrule
TrainCheck pytorch-forum-84911 & -- & \ding{51} & -- & -- & -- & -- \\
TrainCheck transformers-33844 & -- & \ding{51} & -- & -- & -- & -- \\
\midrule
detected & 0/2 & 2/2 & 0/2 & 0/2 & 0/2 & 0/2 \\
false alarm on the fixed version & 0/2 &  & 0/2 & 0/2 & 0/2 & 0/2 \\
\bottomrule
\end{tabular}%
}
\end{table}

\begin{table}[htbp]
\centering\small
\caption{V1 of \tool{} (before) and V2, the version with the post-evaluation
fixes (after), on the same models and operators (seed 0).}
\label{tab:postfix}
\begin{tabular}{lrrrrr}
\toprule
 & & \multicolumn{2}{c}{FAIL (\%)} & \multicolumn{2}{c}{FAIL/WARN (\%)} \\
\cmidrule(lr){3-4}\cmidrule(lr){5-6}
Bug (mutation) & $n$ & before & after & before & after \\
\midrule
noop (control) & 101 & 4 & 3 & 32 & 33 \\
unused\_layer & 103 & 100 & 100 & 100 & 100 \\
detach & 101 & 100 & 100 & 100 & 100 \\
dead\_layer & 101 & 100 & 100 & 100 & 100 \\
batch\_mix & 94 & 69 & 71 & 85 & 88 \\
softmax\_batch & 93 & 84 & 88 & 87 & 91 \\
bn\_batch\_stats & 42 & 100 & 100 & 100 & 100 \\
input\_ignored & 75 & 100 & 100 & 92 & 92 \\
mask\_ignored & 27 & 100 & 100 & 96 & 96 \\
nan\_grad & 101 & 100 & 100 & 100 & 100 \\
dropout\_in\_eval & 101 & 7 & 6 & 70 & 70 \\
explode\_init & 103 & 14 & 12 & 40 & 40 \\
overconfident\_init & 29 & 0 & 0 & 79 & 79 \\
optimizer\_missing & 103 & 100 & 100 & 100 & 100 \\
lr\_zero & 103 & 100 & 100 & 100 & 100 \\
frozen\_breach & 103 & 93 & 96 & 93 & 96 \\
loss\_rewrapped & 103 & 100 & 100 & 100 & 100 \\
loss\_unreduced & 29 & 100 & 100 & 100 & 100 \\
\midrule
all bugs & 1411 & 81 & 82 & 90 & 91 \\
healthy (false alarm) & 103 & 4 & 3 & & \\
\bottomrule
\end{tabular}
\end{table}

\begin{table}[htbp]
\centering\small
\caption{Raw detection rate (\%) per mutation operator on the model zoo (V1): \emph{FAIL}, any
\texttt{FAIL}; \emph{FAIL/WARN}, a \texttt{FAIL} or \texttt{WARN} of a check designed for the bug
(so not a superset of the first column); a baseline, any failure. Last row: runs of the unmodified
models flagged by each tool.}
\label{tab:detzooraw}
\resizebox{\linewidth}{!}{%
\begin{tabular}{lrrrrrrr}
\toprule
Bug (mutation) & $n$ & torchtest2 (FAIL) & torchtest2 (FAIL/WARN) & torchtest (2019) & overfit one batch & smoke test & training monitor \\
\midrule
noop (control) & 101 & 4 & 32 & 18 & 3 & 0 & 23 \\
unused\_layer & 103 & 100 & 100 & 100 & 3 & 0 & 100 \\
detach & 101 & 100 & 100 & 100 & 3 & 0 & 100 \\
dead\_layer & 101 & 100 & 100 & 100 & 11 & 1 & 100 \\
batch\_mix & 94 & 69 & 85 & 17 & 3 & 0 & 22 \\
softmax\_batch & 93 & 84 & 87 & 19 & 2 & 2 & 12 \\
bn\_batch\_stats & 42 & 100 & 100 & 19 & 7 & 0 & 29 \\
input\_ignored & 75 & 100 & 92 & 27 & 12 & 8 & 39 \\
mask\_ignored & 27 & 100 & 96 & 7 & 0 & 0 & 7 \\
nan\_grad & 101 & 100 & 100 & 96 & 100 & 0 & 100 \\
dropout\_in\_eval & 101 & 7 & 70 & 16 & 3 & 0 & 26 \\
explode\_init & 103 & 14 & 40 & 20 & 0 & 0 & 38 \\
overconfident\_init & 29 & 0 & 79 & 7 & 0 & 0 & 66 \\
optimizer\_missing & 103 & 100 & 100 & 100 & 8 & 0 & 100 \\
lr\_zero & 103 & 100 & 100 & 100 & 91 & 0 & 100 \\
frozen\_breach & 103 & 93 & 93 & 17 & 3 & 0 & 23 \\
loss\_rewrapped & 103 & 100 & 100 & 100 & 91 & 0 & 100 \\
loss\_unreduced & 29 & 100 & 100 & 100 & 100 & 100 & 100 \\
\midrule
all bugs & & 81 & 90 & 61 & 26 & 3 & 66 \\
healthy models (false alarm) & & 4 &  & 18 & 2 & 0 & 23 \\
\bottomrule
\end{tabular}%
}
\end{table}

\begin{table}[htbp]
\centering\small
\caption{Mutation-specific detection (\%) per operator on the held-out zoo (V3), defined as in
Table~\ref{tab:detzoo}.}
\label{tab:detheldoutpaired}
\resizebox{\linewidth}{!}{%
\begin{tabular}{lrrrrrrrr}
\toprule
Bug (mutation) & $n$ & tt2 report & tt2 strict & tt2 lenient & torchtest (2019) & monitor & overfit & smoke \\
\midrule
noop (control) & 190 & 0 & 0 & 0 & 0 & 1 & 0 & 0 \\
unused\_layer & 191 & 94 & 97 & 97 & 76 & 87 & 0 & 0 \\
detach & 190 & 94 & 97 & 97 & 76 & 87 & 1 & 1 \\
dead\_layer & 190 & 94 & 97 & 97 & 76 & 87 & 5 & 1 \\
batch\_mix & 188 & 61 & 65 & 86 & 0 & 2 & 0 & 0 \\
softmax\_batch & 189 & 72 & 77 & 98 & 1 & 3 & 0 & 0 \\
bn\_batch\_stats & 95 & 99 & 99 & 99 & 0 & 0 & 0 & 0 \\
input\_ignored & 140 & 98 & 99 & 99 & 3 & 9 & 0 & 0 \\
mask\_ignored & 50 & 84 & 88 & 88 & 0 & 0 & 0 & 0 \\
nan\_grad & 189 & 94 & 99 & 99 & 76 & 87 & 100 & 0 \\
dropout\_in\_eval & 189 & 0 & 0 & 70 & 1 & 2 & 0 & 0 \\
explode\_init & 190 & 16 & 16 & 31 & 3 & 29 & 0 & 0 \\
optimizer\_missing & 190 & 94 & 100 & 100 & 76 & 87 & 2 & 0 \\
lr\_zero & 190 & 94 & 98 & 99 & 76 & 87 & 96 & 0 \\
frozen\_breach & 190 & 94 & 99 & 99 & 0 & 0 & 0 & 0 \\
loss\_rewrapped & 190 & 94 & 99 & 99 & 76 & 87 & 96 & 0 \\
\midrule
all bugs & & 78 & 81 & 90 & 40 & 48 & 22 & 0 \\
\bottomrule
\end{tabular}%
}
\end{table}

\begin{table}[htbp]
\centering\small
\caption{Raw detection rate (\%) on random architectures (V1). Relative to the healthy
counterparts the rates differ by at most one percentage point (Table~\ref{tab:paired}).}
\label{tab:detfuzz}
\resizebox{\linewidth}{!}{%
\begin{tabular}{lrrrrrr}
\toprule
Bug (mutation) & $n$ & torchtest2 (FAIL) & torchtest2 (FAIL/WARN) & torchtest (2019) & overfit one batch & smoke test \\
\midrule
noop (control) & 556 & 0 & 21 & 0 & 0 & 0 \\
unused\_layer & 600 & 100 & 100 & 100 & 0 & 0 \\
detach & 556 & 100 & 100 & 100 & 1 & 0 \\
dead\_layer & 556 & 100 & 100 & 100 & 32 & 0 \\
batch\_mix & 556 & 100 & 100 & 0 & 0 & 0 \\
softmax\_batch & 600 & 100 & 100 & 0 & 1 & 0 \\
bn\_batch\_stats & 155 & 100 & 100 & 0 & 0 & 0 \\
input\_ignored & 600 & 100 & 100 & 0 & 0 & 0 \\
nan\_grad & 556 & 100 & 100 & 100 & 100 & 0 \\
dropout\_in\_eval & 556 & 0 & 100 & 0 & 0 & 0 \\
explode\_init & 600 & 2 & 96 & 2 & 0 & 0 \\
overconfident\_init & 600 & 0 & 93 & 0 & 0 & 0 \\
optimizer\_missing & 600 & 100 & 100 & 100 & 0 & 0 \\
lr\_zero & 600 & 100 & 100 & 100 & 86 & 0 \\
frozen\_breach & 600 & 100 & 100 & 0 & 0 & 0 \\
loss\_rewrapped & 600 & 100 & 100 & 100 & 86 & 0 \\
loss\_unreduced & 600 & 100 & 100 & 100 & 100 & 100 \\
\midrule
all bugs & & 81 & 99 & 52 & 27 & 7 \\
healthy models (false alarm) & & 0 &  & 0 & 0 & 0 \\
\bottomrule
\end{tabular}%
}
\end{table}

\begin{table}[htbp]
\centering\small
\caption{Edge-case study: outcome before and after the fixes.}
\label{tab:edge}
\begin{tabular}{llll}
\toprule
Group & Case & Before & After \\
\midrule
construction & lazy modules & crash & ok \\
precision & fp16 gpu & false FAIL & ok \\
precision & bf16 cpu & ok & ok \\
precision & autocast bf16 & ok & ok \\
precision & float64 & ok & ok \\
wrappers & torch compile & ok & ok \\
wrappers & data parallel & ok & ok \\
parameters & tied weights & ok & ok \\
parameters & weight norm & ok & ok \\
parameters & spectral norm & ok & ok \\
parameters & no parameters & ok & ok \\
parameters & ema teacher no grad & ok & ok \\
outputs & int bool outputs & ok & ok \\
outputs & complex output & false FAIL & ok \\
outputs & scalar output & ok & ok \\
outputs & none in outputs & ok & ok \\
inputs & inplace input & false FAIL & ok \\
inputs & inputs require grad & ok & ok \\
inputs & gpu model cpu inputs & ok & ok \\
inputs & varargs forward & ok & ok \\
inputs & packed lstm unsorted & false FAIL & ok \\
inputs & packed lstm sorted & false FAIL & ok \\
inputs & channels last & ok & ok \\
inputs & batch size one & ok & ok \\
inputs & batchnorm batch one & false FAIL & ok \\
state & counter buffer & ok & ok \\
optimizer & param groups momentum & ok & ok \\
optimizer & warm adamw state & ok & ok \\
optimizer & lbfgs & false FAIL & ok \\
optimizer & sparse embedding & false FAIL & ok \\
training & gradient checkpointing & ok & ok \\
loss & float loss & ok & ok \\
gradcheck & gradcheck tanh & ok & ok \\
gradcheck & gradcheck relu & ok & ok \\
gradcheck & gradcheck int inputs & ok & ok \\
wrappers & jit script & ok & ok \\
outputs & hf tuple outputs & false FAIL & ok \\
state & user hooks & ok & ok \\
inputs & float64 inputs float32 model & ok & ok \\
optimizer & optimizer param group list & ok & ok \\
parameters & module list & ok & ok \\
\bottomrule
\end{tabular}
\end{table}

\begin{table}[htbp]
\centering\small
\caption{Raw detection rate (\%) per mutation operator on the held-out zoo (V3), columns as in
Table~\ref{tab:detzooraw}.}
\label{tab:detheldout}
\resizebox{\linewidth}{!}{%
\begin{tabular}{lrrrrrrr}
\toprule
Bug (mutation) & $n$ & torchtest2 (FAIL) & torchtest2 (FAIL/WARN) & torchtest (2019) & overfit one batch & smoke test & training monitor \\
\midrule
noop (control) & 190 & 6 & 23 & 24 & 0 & 0 & 13 \\
unused\_layer & 191 & 100 & 99 & 100 & 0 & 0 & 100 \\
detach & 190 & 100 & 99 & 100 & 1 & 1 & 100 \\
dead\_layer & 190 & 100 & 99 & 100 & 5 & 1 & 100 \\
batch\_mix & 188 & 66 & 85 & 23 & 0 & 0 & 14 \\
softmax\_batch & 189 & 78 & 95 & 24 & 0 & 0 & 14 \\
bn\_batch\_stats & 95 & 100 & 100 & 26 & 0 & 0 & 17 \\
input\_ignored & 140 & 100 & 100 & 29 & 0 & 0 & 21 \\
mask\_ignored & 50 & 100 & 98 & 16 & 0 & 0 & 10 \\
nan\_grad & 189 & 100 & 99 & 100 & 100 & 0 & 100 \\
dropout\_in\_eval & 189 & 6 & 70 & 25 & 0 & 0 & 14 \\
explode\_init & 190 & 21 & 32 & 25 & 0 & 0 & 41 \\
optimizer\_missing & 190 & 100 & 100 & 100 & 2 & 0 & 100 \\
lr\_zero & 190 & 100 & 99 & 100 & 96 & 0 & 100 \\
frozen\_breach & 190 & 99 & 99 & 24 & 0 & 0 & 13 \\
loss\_rewrapped & 190 & 100 & 99 & 100 & 96 & 0 & 100 \\
\midrule
all bugs & & 83 & 91 & 64 & 22 & 0 & 61 \\
healthy models (false alarm) & & 6 &  & 24 & 0 & 0 & 13 \\
\bottomrule
\end{tabular}%
}
\end{table}

\begin{figure}[htbp]
\centering
\includegraphics[width=0.62\linewidth]{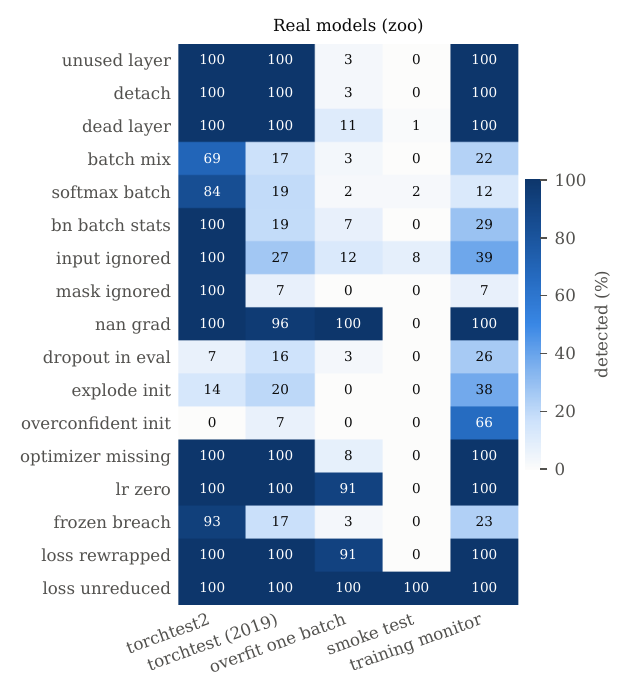}
\caption{Detection rate (\%) per silent bug and tool on the model zoo.}
\label{fig:detzoo}
\end{figure}

\begin{figure}[htbp]
\centering
\includegraphics[width=\linewidth]{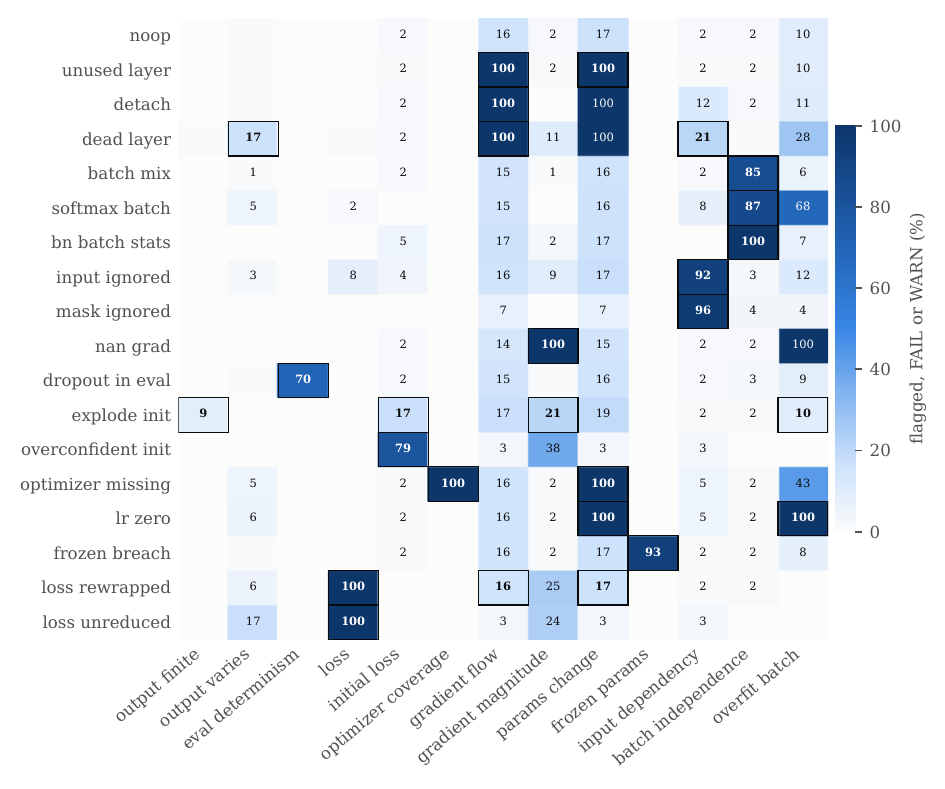}
\caption{Which checks flag each bug (\% of models, FAIL or WARN). Boxed cells: checks designed
for the bug.}
\label{fig:checkszoo}
\end{figure}
\end{document}